\documentclass{lmcs} 

\keywords{guarded fragment, finite model property, probabilistic method}

\usepackage{amsthm}
\usepackage{amsmath}
\usepackage{amssymb}
\usepackage{bm}
\usepackage[linesnumbered,ruled,vlined]{algorithm2e}
\usepackage{hyperref}

\newcommand{\cE}{{\mathcal{E}}}
\newcommand{\cF}{{\mathcal{F}}}

\newcommand{\cO}{{\mathcal{O}}}

\newcommand{\cW}{{\mathcal{W}}}
\newcommand{\cX}{{\mathcal{X}}}

\newcommand{\bbP}{{\mathbb{P}}}

\newcommand{\fA}{{\mathfrak{A}}}
\newcommand{\fB}{{\mathfrak{B}}}

\newcommand{\fM}{{\mathfrak{M}}}
\newcommand{\fN}{{\mathfrak{N}}}

\newcommand{\as}{{\bar{a}}}
\newcommand{\bs}{{\bar{b}}}

\newcommand{\vs}{{\bar{v}}}

\newcommand{\xs}{{\bar{x}}}
\newcommand{\ys}{{\bar{y}}}

\newcommand{\FO}{\mbox{\rm FO}}
\newcommand{\FOt}{\mbox{$\mbox{\rm FO}^2$}}

\newcommand{\GF}{\mbox{$\mbox{\rm GF}$}}

\newcommand{\GFU}{\mbox{$\mbox{\rm GFU}$}}
\newcommand{\CGF}{\mbox{$\mbox{\rm CGF}$}}

\newcommand{\ALC}{$\mathcal{ALC}$}

\newcommand{\TGF}{\mbox{$\mbox{\rm TGF}$}}

\newcommand{\ExpTime}{\textsc{ExpTime}}

\newcommand{\NExpTime}{\textsc{NExpTime}}

\newcommand{\str}[1]{{\mathfrak{#1}}}
\newcommand{\restr}{\!\!\restriction\!\!}

\newcommand{\N}{{\mathbb N}}   
\newcommand{\R}{{\mathbb R}}   

\newcommand{\cutout}[1]{}

\newcommand{\interior}{\operatorname{int}}

\newcommand{\type}[2]{{\rm tp}^{{#1}}[{#2}]}
\newcommand{\deltatype}[2]{\partial{\rm \text{-}tp}^{{#1}}[{#2}]}
\newcommand{\inttype}[2]{{\rm int\text{-}tp}^{{#1}}[{#2}]}

\newcommand{\TTT}{\mbox{\large \boldmath $\tau$}}
\newcommand{\ONE}{\mbox{\large \boldmath $1$}}

\newcommand{\bigmid}{~\big|~}

\newcommand{\Fail}{{\cF ail}}

\newcommand{\FreeVars}{{\rm fv}}

\newcommand{\Cons}{{\rm{Cons}}}
\newcommand{\Rels}{{\rm{Rels}}}

\newcommand{\Width}{{\rm{wd}}}
\newcommand{\Arity}{{\rm{ar}}}

\renewcommand{\phi}{\varphi} 

\let\emptyset\varnothing

\newcommand{\eqdef}{:=}

\renewcommand{\subparagraph}[1]{\smallskip\noindent{\bf #1}}

\theoremstyle{plain} 

\begin{document}

\title{Random Models and the Guarded Fragment}
\titlecomment{{\lsuper*}Funding: Polish National Science Center, grant No.~2021/41/B/ST6/00996.
This article is a journal version of a STACS 2026 conference paper~\cite{Fiuk26a}.}

\author[O.~Fiuk]{Oskar Fiuk}
\address{Institute of Computer Science, University of Wroc\l{}aw, Poland}	
\email{contact.oskarfiuk@gmail.com}  

\hypersetup{
  pdfsubject={Full version of the STACS 2026 paper},
}

\begin{abstract}
  Building on ideas of Gurevich and Shelah for the G\"odel Class, we present a new probabilistic proof of the finite model property for the Guarded Fragment of First-Order Logic. 
  Our proof is conceptually simple and yields the optimal doubly-exponential upper bound on the size of minimal models. 
  We precisely analyse the obtained bound, up to constant factors in the exponents, and construct sentences that enforce models of tightly matching size.
  The probabilistic approach adapts naturally to the Triguarded Fragment, an extension of the Guarded Fragment that also subsumes the Two-Variable Fragment.
  Finally, we derandomise the probabilistic proof by providing an explicit model construction which replaces randomness with deterministic hash functions.
\end{abstract}

\maketitle

\section{Introduction}
\label{sec:intro}

In this work we consider First-Order Logic~(\FO{}) without function symbols of positive arity.
The Guarded Fragment~(\GF{}) is a fragment of \FO{} in which quantifiers are relativised by atomic formulas. 
Syntactically, \GF{} is obtained by restricting quantification to the forms:  
\[
  \forall \xs\,\big(\gamma(\xs,\ys) \rightarrow \psi(\xs,\ys)\big)
  \quad\text{and}\quad
  \exists \xs\,\big(\gamma(\xs,\ys) \wedge \psi(\xs,\ys)\big),
\]
where $\gamma(\xs,\ys)$ is an atomic formula, called a \emph{guard}, mentioning all variables in $\xs$ and $\ys$.  

For example, the following sentence, describing a professor--student scenario, is in \GF{}:
\begin{equation}\label{eqn:example-guarded}
  \forall p,s\,\big(\mathsf{supervises}(p,s) \rightarrow \big(\neg \mathsf{graduate}(s) \wedge \exists t\,\big(\mathsf{prepares}(s,t) \wedge \mathsf{thesis}(t)\big)\big)\big),
\end{equation}
where $\mathsf{supervises}(p,s)$ and $\mathsf{prepares}(s,t)$ serve as guards.
In contrast, the following sentence is not in \GF{}, since the quantifier $\forall p,s$ is not guarded by a single atomic formula:
\begin{equation}\label{eqn:not-example-guarded}
  \forall p,s\;\big(\big(\mathsf{professor}(p) \wedge \mathsf{student}(s)\big) \rightarrow \exists t\,\big(\mathsf{better\text{-}than\text{-}in}(p,s,t) \wedge \mathsf{topic}(t)\big)\big).
\end{equation}

Andréka, van Benthem, and Németi~\cite{ABN98} introduced the Guarded Fragment as a generalisation of modal logic, aiming to transfer its key properties into the richer framework of First-Order Logic.
They established the decidability of satisfiability, and Grädel~\cite{Gra99} later proved the complexity to be $2$-\ExpTime-complete; under bounded number of variables or bounded arity of relation symbols, the complexity drops to \ExpTime-complete.  

The decidability of \GF{} is impressively robust: it is preserved under numerous extensions, including fixed points~\cite{GW99}, transitive or equivalence guards~\cite{KT18,KR21lics}, and (negated) conjunctive queries~\cite{BCO12}. 
Further decidable fragments have been obtained by relaxing the notion of a guard. They include the Loosely, Packed, and Clique-Guarded Fragments~\cite{Ben97,Mar01,Gra99b}, the Guarded Negation Fragment~\cite{BtCS15}, and the Triguarded Fragment~\cite{RS18}.

\subparagraph{Motivations.}
In addition to decidability, a central question in the study of logical fragments is the \emph{finite model property}:
namely, whether every satisfiable sentence of a fragment admits a finite model.
For the Guarded Fragment, the finite model property is known to hold, with a doubly exponential upper bound (in the length of the sentence) on the size of minimal models.

The first proof of the finite model property for \GF{} was given by Grädel~\cite{Gra99}, relying on a deep combinatorial theorem by Herwig~\cite{Her95}. While Grädel’s approach was elegant in its logical formulation—using Herwig’s result as a black box—the underlying construction is technically involved. Moreover, this proof yields only a triply-exponential upper bound on the size of minimal models, which is far from being optimal.
A significant improvement came from Bárány, Gottlob, and Otto~\cite{BGO14}, who established an optimal doubly-exponential bound. Their approach involves analysing finite guarded bisimilar covers of hypergraphs and relational structures, substantially generalising Rosati’s finite chase~\cite{rosati2006finitechase}. In fact, their result extends beyond \GF{}, covering the richer setting where (negations of) conjunctive queries are allowed. A simplified version tailored specifically to \GF{} appears in Pratt-Hartmann’s book~\cite{PH23}, though we believe that even this version remains challenging to follow.

As the existing proofs of the finite model property for \GF{} are unexpectedly difficult, it is natural to ask: \emph{Can we find simpler proofs of this fundamental result?}

\smallskip
A perspective that connects abstract model theory with applied computer science arises in knowledge representation and reasoning, a subfield of artificial intelligence. 
In this context, the Guarded Fragment serves as a logical foundation for decidable reasoning frameworks, subsuming and extending basic description logics (DLs) of the \ALC{} family (for an introduction to DLs, see, e.g.,~\cite{Baader_Horrocks_Lutz_Sattler_2017}).
Here, objects from applications such as databases, knowledge bases, or computer programs are represented as logical structures, while formulas act as a declarative specification language describing their properties. 
Algorithms for satisfiability then become reasoning engines: given a formula, they decide whether an object satisfying the imposed logical constraints exists.

Over the years, a variety of algorithms solving satisfiability for the Guarded Fragment have been developed, ranging from purely theoretical decision procedures~(e.g.,~\cite{Gra99}) to practically implementable methods based on resolution, saturation, or tableau~(e.g.,~\cite{NIVELLE200321,Kaz06,hirsch2002guarded}). 
A key aspect is that many of these algorithms not only decide satisfiability but also produce a finite combinatorial object: a certificate of formula consistency. 

Our interest lies in the subsequent step: \emph{How to turn such certificates into explicit finite models?}
Since smaller models are typically more useful—both for computational efficiency and for practical interpretability—the central challenge is to generate finite models as small as possible:
not only close to the theoretical doubly exponential bounds, typically considered up to polynomial slack in the second-level exponent, but actually sharpened to within constant factors and detailed structural parameters of sentences.

\subparagraph{Main Contribution.}
In this work, we give a new proof of the finite model property for \GF{},
yielding the optimal doubly exponential bound on the size of minimal models.
To the best of our knowledge, no earlier proof exhibits comparable simplicity and self-containment.

We employ a probabilistic approach inspired by Gurevich and Shelah’s proof of the finite model property for the Gödel Class~\cite{GS83}. 
The central idea is the following: given a formula $\phi$ and a (possibly infinite) model $\str{A}$ with $\str{A}\models\phi$, we define a random procedure that generates a finite structure $\str{B}_n$ with domain of size $n \in \N$. 
We then prove that, once $n$ exceeds a certain threshold depending only on $\phi$, the probability that $\str{B}_n \models \phi$ becomes strictly positive. 
Consequently, some finite $\str{B}_n$ must be a model of $\phi$, yielding the finite model property with an upper bound on the minimal model size that matches this threshold.

Applying this probabilistic method to the Guarded Fragment, we establish our main result:
every satisfiable \GF{}-sentence $\phi$ has a finite model whose domain size is
\(
  2^{2^{\,\cO(|\phi|\cdot\log|\phi|)}}.
\)

In this work, we assume that $|\phi|$ measures the \emph{uniform length} of $\phi$:
formulas are viewed as words over an infinite alphabet consisting of parentheses, logical connectives, quantifiers, variables, relation symbols, and constants, with each symbol contributing $1$ to the length.\footnote{The uniform length differs from the bit-length of formulas (finite alphabet), in which $n$ distinct variables or symbols requires $\Theta(n\cdot \log n)$ bits.}

To witness the tightness of our upper bound, we construct a family of sentences $(\phi_n)_{n \in \N}$ whose minimal models have domains of size at least
\(
  2^{2^{\,\Omega(|\phi_n|\cdot\log|\phi_n|)}}.
\)
It is an improvement upon earlier approaches for enforcing large models in \GF{}, as they only achieve domains of size 
\(
  2^{2^{\,\Omega(\sqrt{|\phi_n|})}}
\) 
(cf.~\cite{Gra99}).
Consequently, our probabilistic model construction yields an essentially optimal upper bound, up to constant factors in the second-level exponent.

\begin{thm}\label{theorem:main}
  There exist universal constants $0 < C_{\mathrm{lb}} < C_{\mathrm{ub}}$ such that the following holds.
  \begin{enumerate}
    \item\label{theorem:main:item1} Every satisfiable \GF-sentence $\varphi$ has a finite model whose domain size is at most
    \[
      2^{2^{\,C_{\mathrm{ub}}\cdot |\varphi|\cdot\log|\varphi|}}.
    \]
    \item\label{theorem:main:item2} For every $n \in \N$, there exists a satisfiable \GF-sentence $\varphi_n$
    with \(|\varphi_n| \ge n\) such that any model of $\varphi_n$ must have domain size at least
    \[
      2^{2^{\,C_{\mathrm{lb}}\cdot |\varphi_n|\cdot\log|\varphi_n|}}.
    \]
  \end{enumerate}
\end{thm}

The upper bound of Theorem~\ref{theorem:main} is clean and elegant, but it is stated solely in terms of the length of~$\phi$ and an unspecified constant in the second-level exponent.
We complement this result by deriving a more precise bound on the size of minimal models.
For a concise formulation, we express it in terms of the number of \emph{(atomic) $k$-types}—that is, maximal consistent configurations of literals over $k$ variables (see Section~\ref{sec:prelim} for a formal definition).

Given a sentence $\phi$, its \emph{induced signature} $\sigma$ is the set of all relation and constant symbols occurring in~$\phi$.  
The \emph{width} of $\sigma$, denoted $\Width(\sigma)$, is the maximum arity of any relation symbol in~$\sigma$.
We define the \emph{expanded normal-form signature} $\sigma_{\mathrm{nf}}$ from $\sigma$ as follows:
for every subformula $\chi$ of $\phi$ that begins with a maximal block of quantifiers and which is not a sentence (i.e., $\chi$ has free variables), we introduce a fresh relation symbol $R_\chi$ whose arity equals the number of free variables of $\chi$ (which is at most $\Width(\sigma)$, since $\chi$ is required to be guarded).  
Note that the construction of $\sigma_{\mathrm{nf}}$ preserves width, i.e., $\Width(\sigma_{\mathrm{nf}}) = \Width(\sigma)$,
does not introduce new constants, and keeps the overall size of $\sigma_{\mathrm{nf}}$ linear in $|\phi|$.

\begin{thm}\label{theorem:precise-bound}
There exist sequences $(C_t)_{t\in\mathbb{N}}$ and $(\varepsilon_t)_{t\in\mathbb{N}}$ with $C_t > 0$, $\varepsilon_t \in (0,1/e)$, and $\varepsilon_t \to 0$
such that the following holds for every $t \in \mathbb{N}$.

Let $\phi$ be any satisfiable \GF-sentence, and let $\sigma_{\mathrm{nf}}$ be its expanded normal-form signature.
Denote $k = \Width(\sigma_{\mathrm{nf}})$.
If $k \ge t \ge 2$, then $\phi$ has a finite model whose domain size is at most
\[
  C_t \cdot \bigl|\TTT^{\sigma_{\mathrm{nf}}}_{k}\bigr|^{\,1 - 1/e + \varepsilon_t},
\]
where $\TTT^{\sigma_{\mathrm{nf}}}_{k}$ is the set of all $k$-types over $\sigma_{\mathrm{nf}}$.
\end{thm}

The exponent in Theorem~\ref{theorem:precise-bound} converges to $1 - 1/e \approx 0.63$ as $t \rightarrow \infty$.
Informally, Theorem~\ref{theorem:precise-bound} can thus be read as stating that every satisfiable \GF{}-sentence admits a finite model of domain size proportional to
\( \bigl|\TTT^{\sigma_{\mathrm{nf}}}_{k}\bigr|^{\,0.63}, \)
provided that the width $k = \Width(\sigma_{\mathrm{nf}})$ is sufficiently large.

\smallskip
At first sight, working with the expanded normal-form signature rather than the induced one may seem unintuitive.
However, this choice is in fact a consequence of the semantics of guarded sentences:
a quantified subformula $\chi(\ys)$ with free variables $\ys$ naturally defines a relation in a structure $\fA$, namely
\(R^{\fA}_{\chi} = \{\, \as \mid \fA \models \chi(\as) \,\}.\)
To simplify the quantifier structure of a sentence, a standard technique is to transform it into a suitable \emph{normal form} (see Section~\ref{sec:background} for a definition).
The normal-form reduction makes such implicitly defined relations explicit by introducing corresponding relation symbols.
As a result, it distinguishes elements that are locally—i.e., at the level of atomic types—indistinguishable in the induced signature,
but differ in their satisfaction of quantified subformulas.

\begin{exa}\label{example:normal-form-signature}
  To illustrate, consider the following (slightly abstract) \GF-sentence:
  \begin{equation*}\label{eqn:normal-form-signature}
    \exists x\;\big(U(x) \wedge \exists y,z\;\big(P(x,y,z) \wedge \neg U(y)\big)\big)
    \;\wedge\;
    \exists u\;\big(U(u) \wedge \forall v,w\;\big(P(u,v,w) \rightarrow U(v)\big)\big).
  \end{equation*}
  This sentence is satisfied in a model $\fA$ with domain $\{1,2,3\}$, where
  $U^{\fA} = \{1,3\}$ and $P^{\fA} = \{(1,2,3)\}$.  
  The variables $x$, $y$, $z$ are witnessed by the elements $1$, $2$, $3$; and the variable $u$ by element~$3$.
  Notice, however, that in the induced signature $\sigma = \{ P,\, U \}$, elements $1$ and $3$ are indistinguishable at the level of $1$-types:
  both satisfy only the unary predicate $U$.  
  
  Let $\mu$ and $\nu$ denote the subformulas that begin with quantifiers $\exists y,z$ and $\forall v,w$, respectively.
  The expanded normal-form signature introduces two additional predicates $R_{\mu}$ and $R_{\nu}$,
  which record the different roles of $1$ and $3$ in the model: $R^{\str{A}}_{\mu} = \{ 1 \}$ and $R^{\str{A}}_{\nu} = \{ 2, 3 \}$.
  
  Finally, note that the subformulas starting with $\exists x$ or $\exists u$ are subsentences;
  and those with $\exists z$ or $\forall w$ do not begin with a maximal block of quantifiers.
  Hence we do not introduce fresh relation symbols for them.
  Thus the resulting signature is $\sigma_{\mathrm{nf}} = \{ P,\, U,\, R_{\mu},\, R_{\nu} \}$.
\end{exa}

\subparagraph{Triguarded Fragment.}
The Guarded Fragment~(\GF{}) and the Two-Variable Fragment~(\FOt{}) are among the most prominent fragments of First-Order Logic.
Both capture a wide range of modal and description logics and are decidable, yet they differ substantially and are incomparable in expressive power.
In particular, \GF{} cannot express certain basic properties, expressible in \FOt{}, such as
\(
  \forall s,d\;\big(\big(\mathsf{student}(s) \wedge \mathsf{dean}(d)\big) \rightarrow \mathsf{knows}(s,d)\big).
\)

To unify these two perspectives, Rudolph and {\v{S}}imkus~\cite{RS18} introduced the \emph{Triguarded Fragment} (\TGF{}), building on related ideas developed earlier in~\cite{Kaz06,BMP17}.
The key idea of \TGF{} is to relax the quantification restrictions of \GF{}: formulas with at most two free variables may be quantified freely, while the guardedness requirement is retained for formulas with three or more free variables (hence the name “tri-guarded”).  
This way, \TGF{} subsumes both \FOt{} and \GF{}, while also capturing properties beyond their reach.  
For instance, in formula~\eqref{eqn:not-example-guarded}, the quantifier $\forall p,s$ is admissible in \TGF{} because the quantified subformula has only two free variables, $p$ and $s$, whereas the quantifier $\exists t$ is required to be guarded, and indeed is by the atom $\mathsf{better\text{-}than\text{-}in}(p,s,t)$.  
Hence~\eqref{eqn:not-example-guarded} belongs to \TGF{}.

An important distinction concerns the role of equality.  
While both \FOt{} and \GF{} remain decidable in the presence of equality, the satisfiability problem for \TGF{} with equality is undecidable.  
The reason is that \TGF{} is expressive enough to encode the G\"odel Class (i.e., the prefix class $\forall\forall\exists^*$), which was shown to be undecidable in the presence of equality by Goldfarb~\cite{Gol84}.
Excluding equality suffices to restore decidability of satisfiability, with complexity being $2$-\ExpTime{}-complete without constants and $2$-\NExpTime{}-complete when constants are allowed~\cite{RS18}.

Returning to the finite model property, 
Kiero\'nski and Rudolph~\cite{KR21lics} proved that the equality-free fragment of \TGF{} has the finite model property, with an optimal doubly exponential bound on minimal model size.  
Their proof, however, is technically intricate: it uses the finite model property for \GF{} as a black box and adds a further combinatorial construction that carefully glues several structures into a single model.  

In contrast, our probabilistic approach to \GF{} extends seamlessly to \TGF{}, yielding a considerably simpler proof of the finite model property for this broader fragment. Moreover, it also achieves the optimal doubly-exponential upper bound on minimal model size.

\begin{thm}\label{theorem:TGF-main}
  There exists a universal constant $C > 0$ such that the following holds.
  Let $\varphi$ be a satisfiable equality-free sentence of \TGF{}.
  Then $\varphi$ has a finite model whose domain size is at most
  \[
    2^{2^{\,C \cdot |\varphi|\cdot\log|\varphi|}}.
  \]
\end{thm}

\subparagraph{Derandomisation.}
As mentioned earlier, the satisfiability problem for the Guarded Fragment is $2$-\ExpTime-complete~\cite{Gra99}.
Yet, although the probabilistic construction shows that every satisfiable sentence has a model of doubly exponential size, this alone does not imply that such a model can be constructed in time matching the decision complexity.
Indeed, the number of structures of doubly exponential size is triply exponential.

From a practical point of view, this gap is minor: our randomised construction succeeds with probability at least \(1/2\) while sampling structures over a domain of doubly exponential size.  
Nevertheless, from a theoretical standpoint it leaves a conceptual separation between reasoning and model building: deterministic vs. randomised time complexity.

We close this gap by providing a constructive, fully deterministic version of Theorem~\ref{theorem:main}.  
The central idea is to replace the random choices by carefully selected deterministic ones, 
using families of hash functions that mimic the statistical properties of true randomness.  
We then prove that these deterministic choices always produce a valid model.  

In our formulation, we rely on the notion of a \emph{witness of satisfiability}:
a finite combinatorial object—namely, a set of $k$-types satisfying specific closure and consistency properties—that certifies the satisfiability of a sentence.
Its definition is deferred to Section~\ref{sec:background}; 
for the present discussion it suffices to know that such a witness can be computed from a satisfiable sentence in $2$-\ExpTime{}~\cite{Gra99,PH23}.  
On this basis we obtain the following constructive guarantee:

\begin{prop}\label{prop:constructive}
  There exists a deterministic algorithm with the following property:  
  given a witness of satisfiability $\cW$ for a \GF{}-sentence $\phi$,  
  the algorithm constructs a structure $\fB$ with domain of size $n \in \N$ such that $\fB \models \phi$,  
  where the parameter $n$ satisfies $n = 2^{2^{\,\cO(|\phi|\cdot\log|\phi|)}}$.
  The algorithm runs in doubly exponential time in the length of~$\phi$.
\end{prop}

\subparagraph{Related Work.}
Beyond Gurevich and Shelah's approach for the Gödel class~\cite{GS83},
the probabilistic method has been employed in several other proofs of the finite model property.

One of the earliest and most influential applications of probabilistic techniques in logic
is Fagin’s proof of the $0$--$1$ law for First-Order Logic~\cite{Fagin76}.
A notable consequence of Fagin’s argument is that any finite subset of the theory of the Rado graph admits a finite model.

Goldfarb, Gurevich, and Shelah subsequently extended the approach of the latter two authors
to the so-called subminimal Gödel class with identity~\cite{GGS84}.
Goldfarb later developed probabilistic proofs for additional decidable fragments,
namely a solvable Skolem class~\cite{Gol93} and the Maslov class~\cite{Gol89}.
For recent applications of the probabilistic method in establishing decidability and the finite model property, see~\cite{FK-lmcs25,FKM24}.

\subparagraph{Technical Overview.}
We begin in Section~\ref{sec:prelim} by introducing the basic notation and conventions used throughout the paper.  
Section~\ref{sec:background} provides the necessary background on \GF{}, including its syntax, the normal form, and a satisfiability criterion.  
Building on this, Section~\ref{sec:random} develops probabilistic constructions of finite models for \GF{}.  
In Section~\ref{sec:tight}, we construct sentences enforcing models that tightly match the upper bound established in Section~\ref{sec:random}.  
Section~\ref{sec:beyond} demonstrates that our methods extend naturally to \TGF{}.  
In Section~\ref{sec:algebra}, we give a deterministic procedure for constructing finite models.
Finally, Section~\ref{sec:conclusions} concludes the paper and discusses directions for future research.  

This is the full version of a paper that appears in the Proceedings of STACS 2026.
In addition to providing technical details omitted from the conference version, it also includes further results, remarks, and examples.

\section{Preliminaries and Notation}
\label{sec:prelim}

We denote the set of natural numbers including $0$ by $\N$. For $k \in \N$, the notation $[k]$ stands for the set $\{1, \dots, k\}$, with the convention that $[0] = \emptyset$. More generally, we use interval notation $[a, b] \subseteq \N$ to denote the set $\{a, a+1, \dots, b\}$ whenever $a \leq b$, and the empty set $\emptyset$ whenever $a > b$.
For a set $S$, we denote by $2^S$ the powerset of $S$, and by $\binom{S}{k}$ the set of all $k$-subsets of $S$.
If $S \subseteq \N$, we also use $\binom{S}{k}$ for the set $\{\langle a_1,\dots,a_k \rangle \in S^k \mid a_1 < \dots < a_k \}$.

\subparagraph{First-Order Logic.}
We assume general familiarity with First-Order Logic ($\FO$).  
The logical symbols are $=, \bot, \top, \vee, \wedge, \neg, \rightarrow, \leftrightarrow$, and the quantifiers $\forall, \exists$.  
Formulas may also use non-logical symbols: relation symbols of arbitrary arity (from a countably infinite set), constant symbols (also from a countably infinite set), and variables (again countably many).  
We do not allow function symbols of positive arity.
The \emph{length} of a formula $\phi$, denoted $|\phi|$, is defined as the total number of symbols it contains, where each occurrence of a symbol---be it a variable, relation symbol, or constant---contributes $1$.
We use $\FreeVars(\phi)$ to denote the set of \emph{free} variables of $\phi$.

A \emph{signature} $\sigma$ is a finite set of symbols, partitioned as $\sigma = \Rels \uplus \Cons$, where $\Rels$ is the set of relation symbols and $\Cons$ is the set of constant symbols.
We require $\Rels \neq \emptyset$.
Every relation symbol $R \in \Rels$ comes with associated arity, denoted $\Arity(R)$.
We require $\Arity(R) \ge 1$.
The \emph{width} of $\sigma$, denoted $\Width(\sigma)$, is the maximum arity of any symbol in $\Rels$.
The \emph{signature of a formula} is the finite set of relation and constant symbols that appear in the formula.

We use Fraktur letters such as $\str{A}, \str{B}, \dots$ to denote structures, and the corresponding Roman letters $A, B, \dots$ for their domains.
A $\sigma$-structure $\str{A}$ is a structure that interprets the symbols in $\sigma$: a relation symbol $R$ as a relation $R^{\str{A}}\subseteq A^k$ with $k$ denoting the arity of $R$; and a constant symbol $c$ as an element $c^{\str{A}} \in A$.
If $B \subseteq A$, we write $\str{A} ~\restr~ B$ for the \emph{restriction} of $\str{A}$ to the subdomain $B$. Note that $B$ must include all interpretations of constant symbols to remain a $\sigma$-structure.
The structure $\fA$ is \emph{empty} if it contains no facts, i.e., $R^{\fA} = \emptyset$ for all relation symbols $R$; however, we do not insist (and even forbid) that the domain is the empty set $\emptyset$. 
The \emph{size} of a structure is the cardinality of its domain.

With respect to satisfiability, equality between constants can be eliminated by a straightforward reduction:
if a formula has a model $\fA$ satisfying $c^{\fA}_1 = c^{\fA}_2$, then $c_2$ can be replaced by $c_1$ throughout the formula, without affecting satisfiability.
Thus we work under the \emph{standard name assumption} requiring that constants are interpreted in structures by themselves.
Given a $\sigma$-structure $\fA$, we partition its domain $A$ as $A_0 \uplus \Cons{}$, where $A_0$ and $\Cons$ are disjoint sets of \emph{unnamed} and, respectively, \emph{named} elements.

\subparagraph{Types.}
Fix a signature $\sigma = \Rels \uplus \Cons$.  
For $k \in \N$, let $\operatorname{Lit}_k(\sigma)$ denote the set of all literals over relation and constant symbols from $\sigma$ and variables $x_1,\dots,x_k$; literals involving ``$=$'' are not included in this set.

For $k \in \N$, a \emph{$k$-type} over $\sigma$ is a maximal consistent subset of $\operatorname{Lit}_k(\sigma)$,
that is, a set including precisely one of $\gamma$, $\neg \gamma$ for every atom $\gamma$ over $\sigma$ and $x_1,\dots,x_k$. 
A \emph{type} is simply a $k$-type for some $k \in \N$.  
We write $\TTT^{\sigma}_k$ for the set of all $k$-types over $\sigma$, and $\TTT^{\sigma} \eqdef \bigcup_{k \in \N}\TTT^{\sigma}_k$.

\begin{clm}\label{claim:number-of-types}
  For every $k \in \N$, we have
  \[ \big|\TTT^{\sigma}_k\big| = \prod_{R \in \Rels} 2^{(k+|\Cons|)^{\Arity(R)}}
  \quad\text{and thus}\quad
  2^{k^{\Width(\sigma)}} \le \big|\TTT^{\sigma}_k\big| \le 2^{|\Rels| \cdot (k + |\Cons|)^{\Width(\sigma)}}. \]
\end{clm}

We denote by $\tau_{\mathrm{all\text{-}neg}}$ the type consisting exclusively of negative literals (the value of $k$ will be clear from the context).

If $k \le \ell$, we write $\tau_2 \models \tau_1$ to indicate that a $k$-type $\tau_1$ is contained in an $\ell$-type $\tau_2$.  

For $\tau \in \TTT^{\sigma}_k$, we define:  
\( \partial\tau \eqdef \{ \gamma \in \tau \mid \FreeVars(\gamma) = \{x_1,\dots,x_k\} \},\) 
and \(\interior(\tau) \eqdef \tau \setminus \partial\tau\),
that is, the \emph{boundary} and \emph{interior} of $\tau$.  
The set of all \emph{boundary $k$-types} is $\partial\TTT^{\sigma}_k \eqdef \{ \partial\tau \mid \tau \in \TTT^{\sigma}_k \}$.

A $k$-type $\tau$ is called \emph{guarded} if either $k \le 1$ or $\tau$ contains a positive literal $\gamma$ with  
\( \FreeVars(\gamma) = \{x_1,\dots,x_k\}, \) i.e., $\gamma \in \partial\tau$. 
Note that necessarily $k \le \Width(\sigma)$.

Given a $k$-tuple of distinct unnamed elements $\langle a_1,\dots,a_k \rangle$ of a $\sigma$-structure $\fA$, we write 
\( \type{\fA}{a_1,\dots,a_k} \) 
for the unique $k$-type \emph{realised} by $\langle a_1,\dots,a_k \rangle$ in $\fA$.  
Formally, $\type{\fA}{a_1,\dots,a_k}$ is the set of literals $\gamma \in \operatorname{Lit}_k(\sigma)$ such that $\fA,f \models \gamma$, where $f$ is the assignment that sends $x_i \mapsto a_i$ for every $i \in [k]$.
The collection of all $k$-types realised in $\fA$ is denoted
\[
  \TTT^{\fA}_k \eqdef \{ \, \type{\fA}{a_1,\dots,a_k} \mid a_1,\dots,a_k \in A \setminus \Cons \text{ are pairwise distinct} \, \}.
\]

It is convenient to view $k$-types themselves as $\sigma$-structures over the canonical domain $\{x_1,\dots,x_k\} \cup \Cons$.  
This allows us to use structure-like notation; for instance, if $\tau \in \TTT^{\sigma}_3$, we may write 
\( \type{\tau}{x_1,x_3} \) 
for the $2$-type induced by $\tau$ on the variables $x_1$ and $x_3$.

\begin{rem}
Our definition of $k$-types deviates slightly from the standard one in two aspects.  
First, we only consider tuples of unnamed elements.  
Second, we restrict attention to tuples of pairwise distinct elements.  
This choice is convenient: to fully describe a structure over a domain $A_0 \uplus \Cons$, it suffices to specify for each tuple of distinct elements of $A_0$ the corresponding realised $k$-type, with each unordered tuple being considered precisely once.
Naturally, the $k$-types have to be assigned consistently with each other.
Since types include literals involving constants, the facts involving constants
are defined as well.
\end{rem}

\section{Technical Background on the Guarded Fragment}
\label{sec:background}

In this section we formally introduce the syntax of the Guarded Fragment, together with the normal form and a satisfiability criterion. 
The material presented here is mostly a direct adaptation of Gr\"adel's work~\cite{Gra99} and is included primarily for the reader’s convenience.
In particular, no claims of novelty are made here.
For a comprehensive introduction to the Guarded Fragment, see also Chapter~4 in Pratt-Hartmann’s monograph~\cite{PH23}.

\subparagraph{Syntax of GF.}
We formalise the syntax of the Guarded Fragment in Definition~\ref{definition:GF-syntax}.

\begin{defi}\label{definition:GF-syntax}
  The \emph{Guarded Fragment} (\GF{}) is the set of formulas in First-Order Logic generated by the following rules:
  \begin{enumerate}[label=(\roman*)]
    \item\label{GF-rule1} Every atomic formula belongs to \GF{}.
    \item\label{GF-rule2} \GF{} is closed under Boolean connectives.
    \item\label{GF-rule3} Let $\xs$ be a tuple of variables, let $\psi$ be a formula in \GF{}, and let $\gamma$ be an atomic formula.
    If $\FreeVars(\psi) \subseteq \FreeVars(\gamma)$ and $\xs \subseteq \FreeVars(\gamma)$,  
    then both $\exists\xs\,(\gamma \wedge \psi)$ and $\forall \xs\,(\gamma \rightarrow \psi)$ belong to \GF{}.
  \end{enumerate}
  The atom $\gamma$ in rule~\ref{GF-rule3} is called a \emph{guard}.
\end{defi}

Note that equality may serve as a guard: the formula $\forall x~(x = x \rightarrow \psi)$ is in \GF{} whenever $\psi \in \GF{}$ and $\FreeVars(\psi) \subseteq \{x\}$.
This allows for free quantification over individual elements.

\subparagraph{Normal Form.}
Let $\phi$ be a sentence in \GF{}.
Then $\phi$ is in \emph{normal form} if it is a conjunction of a finite number of guarded \emph{existential}, guarded \emph{universal}, and guarded \emph{Skolem} sentences:
\[
  \bigwedge_{t} \exists \xs\;\big(\alpha_t(\xs) \wedge \psi_t(\xs)\big) \; \wedge \;
  \bigwedge_{t} \forall \xs\;\big(\alpha_t(\xs) \rightarrow \psi_t(\xs)\big) \; \wedge \;
  \bigwedge_{t} \forall \xs\;\big(\alpha_t(\xs) \rightarrow \exists \ys\;\big(\beta_t(\xs,\ys) \wedge \psi_t(\xs,\ys)\big)\big),
\]
where $\xs$ and $\ys$ are disjoint tuples of distinct variables, $\alpha_t(\xs)$ and $\beta_t(\xs,\ys)$ are guards, and $\psi_t$ are quantifier-free formulas.
The notation $\psi(\xs)$ only highlights that the set of free variables of $\psi$ is contained in the tuple $\xs$.
However, it does not imply that all of them are actually used by $\psi$. 
In particular, when $\psi$ is an atom, the variables can occur in any order and with repetitions.
Also the actual sets of free variables of $\alpha_t$, $\beta_t$, and $\psi_t$ can vary with~$t$.
Nevertheless, we require that each conjunct is properly guarded.  
For every relevant index~$t$, the following must hold:  
if~$t$ corresponds to an existential or universal conjunct, then  
$\FreeVars(\psi_t) \subseteq \FreeVars(\alpha_t) = \xs$;  
if~$t$ corresponds to a Skolem conjunct, then  
$\FreeVars(\alpha_t) = \xs$ and  
$\FreeVars(\psi_t) \cup \ys \subseteq \FreeVars(\beta_t) \subseteq \xs \cup \ys$,  
where in both cases the tuples $\xs$ and $\ys$ may depend on $t$.

\smallskip  
Lemma~\ref{lemma:GF-normal-form} reduces the finite model property for \GF{} to the normal-form case.
Moreover, any upper bound on minimal model size that depends only on the length and signature, and is proved for normal-form sentences, naturally carries over to arbitrary \GF{}-sentences.

\begin{lem}\label{lemma:GF-normal-form}
Let $\phi$ be a \GF{}-sentence over a signature $\sigma$.
Then there exists a normal-form \GF{}-sentence $\phi_{\mathrm{nf}}$
over an expanded signature $\sigma_{\mathrm{nf}} \supseteq \sigma$
such that the following conditions hold.
\begin{enumerate}
  \item The sentences $\phi$ and $\phi_{\mathrm{nf}}$ are equisatisfiable.
  \item If $\fB$ is a $\sigma_{\mathrm{nf}}$-structure with $\fB \models \phi_{\mathrm{nf}}$, 
  then the $\sigma$-reduct of $\fB$ is a model of $\phi$.
  \item $|\phi_{\mathrm{nf}}| \le C \cdot |\phi|$ for some fixed constant $C > 0$.
  \item The expanded signature $\sigma_{\mathrm{nf}}$ is obtained from $\sigma$ as follows:
  for every subformula $\chi$ of $\phi$ that begins with a maximal block of quantifiers
  and is not a sentence (i.e., has free variables),
  introduce a fresh relation symbol $R_\chi$ whose arity equals the number of free variables of $\chi$
  (which is at most $\Width(\sigma)$, since $\chi$ is required to be guarded).
\end{enumerate}
\end{lem}
\begin{proof}
  Let $\phi$ be a $\GF$-sentence.
  If $\phi$ is unsatisfiable, then the lemma holds trivially, e.g., take $\phi_{\mathrm{nf}} \eqdef \exists x\; (x = x \wedge \bot)$.
  Now, assume that $\phi$ has a (possibly infinite) model $\str{A}$. 
  We describe a method for rewriting $\phi$ into normal form.

  W.l.o.g.~we may assume that all quantifiers in $\phi$ are existential, as
  $\forall \xs\;\big(\gamma(\xs,\ys) \rightarrow \psi(\xs,\ys)\big)$ is equivalent to $\neg \exists \xs\;\big(\gamma(\xs,\ys) \wedge \neg \psi(\xs,\ys)\big)$.  
  
  Consider a subformula $\chi(\ys)$ of $\phi$ in the form
  \(
    \chi(\ys) = \exists \xs\;\big( \gamma(\xs,\ys) \wedge \psi(\xs,\ys) \big),
  \)
  chosen to be as deep as possible (i.e., not containing further quantifiers inside $\psi$).  
  Here we consider only subformulas starting with a maximal block of consecutive quantifiers (consult Example~\ref{example:normal-form-signature}).

  First, if $\chi(\ys)$ is a subsentence, i.e., $\ys = \emptyset$,
  then replace the occurrence of $\chi$ with $\top$ if $\str{A}\models \chi$; and with $\bot$ if $\str{A}\models \neg\chi$.
  Then append to $\phi$ respectively either
  \[
    \exists \xs\;\big(\gamma(\xs) \wedge \psi(\xs)\big)
    \quad \text{or} \quad
    \forall \xs\;\big(\gamma(\xs) \rightarrow \neg \psi(\xs)\big).
  \]

  Otherwise, i.e., when $\ys \neq \emptyset$,
  introduce a fresh relation symbol $R_\chi$ of arity $|\ys|$.
  Replace the occurrence of $\chi(\ys)$ in $\phi$ with $R_\chi(\ys)$, and conjoin the following two sentences:
  \[
    \forall \ys\;\big(R_\chi(\ys) \rightarrow \exists \xs\;\big(\gamma(\xs,\ys) \wedge \psi(\xs,\ys)\big)\big)
    \quad \text{and} \quad
    \forall \xs,\ys\;\big(\gamma(\xs,\ys) \rightarrow \big(\psi(\xs,\ys) \rightarrow R_\chi(\ys)\big)\big).
  \]
  The latter sentence is equivalent to $\forall \ys\;\big(\exists \xs\;\big(\gamma(\xs,\ys) \wedge \psi(\xs,\ys)\big) \rightarrow R_\chi(\ys)\big)$.
  Intuitively, $R_\chi$ now acts as a placeholder for $\chi(\ys)$, while the above sentences guarantee equivalence.

  Repeating this process iteratively yields a sentence $\phi_{\mathrm{nf}}$ in normal form over an expanded signature $\sigma_{\mathrm{nf}}$. The required properties of $\phi_{\mathrm{nf}}$ are readily verified. 
\end{proof}

\subparagraph{Satisfiability Criterion.}
We now formulate a sufficient criterion for the satisfiability of sentences in \GF{}.  
This allows us to modularise the argument and clearly separate the general properties of guarded logic from the new components of our proof.

Let $\sigma = \Rels \uplus \Cons$ be a signature.  
A set of types $\TTT \subseteq \TTT^{\sigma}$ is said to be \emph{closed} if the following holds:  
for every $\tau \in \TTT$, which is a $k$-type for some $k$, for every $\ell \in [0,k]$, and for every choice of distinct indices $i_1,\dots,i_\ell \in [k]$, the type
\(
  \type{\tau}{x_{i_1},\dots,x_{i_\ell}}
\)
also belongs to $\TTT$.  
Further, the set $\TTT$ is said to be \emph{consistent} if it contains a unique $0$-type.

A natural candidate for a closed and consistent set of $k$-types is the collection of all $k$-types with $k \le \Width(\sigma)$ that are realised in some structure.

\begin{defi}
  Let $\sigma = \Rels \uplus \Cons$ be a signature, and let \( \TTT = \bigcup_{k=0}^{\Width(\sigma)} \TTT_k \)
  be a family of sets of $k$-types over $\sigma$, where $\TTT_k \subseteq \TTT_k^{\sigma}$ for each $k \in [0,\Width(\sigma)]$.  
  Let $\fA$ be a $\sigma$-structure with domain $A$, and let $A_k = \{\langle a_1,\dots,a_k \rangle \in (A \setminus \Cons)^k \mid a_1,\dots,a_k \; \text{are pairwise distinct} \}$.
  \begin{enumerate}[label=(\roman*)]
    \item We say that $\fA$ is \emph{$\TTT$-guarded} if, for every $k \in [0,\Width(\sigma)]$ and every $\as \in A_k$, whenever the $k$-type $\type{\fA}{\as}$ is guarded, it holds that $\type{\fA}{\as} \in \TTT_k$.
    \item We say that $\fA$ has the \emph{$\TTT$-extension property} if, for every $k \in [0,\Width(\sigma){-}1]$ and every pair $(\tau_1,\tau_2) \in \TTT_k \times \TTT_{k+1}$ such that $\tau_2 \models \tau_1$, the following holds:  
    whenever $\as \in A_k$ satisfies $\type{\fA}{\as} = \tau_1$, there exists an element $b \in A \setminus (\as \cup \Cons)$ such that $\type{\fA}{\as,b} = \tau_2$.
  \end{enumerate}
\end{defi}

\begin{lem}\label{lemma:witness-of-satisfiability}
  Let $\phi$ be a normal-form \GF{}-sentence over a signature $\sigma = \Rels \uplus \Cons$.
  If $\phi$ is satisfiable,
  then there exists a closed and consistent family of sets of $k$-types $\TTT^* = \bigcup_{k=0}^{\Width(\sigma)} \TTT^*_k$, where $\TTT^*_k \subseteq \TTT^{\sigma}_k$ for each $k \in [0,\Width(\sigma)]$, such that the following holds:
  whenever a $\sigma$-structure $\fB$ is $\TTT^{*}$-guarded and satisfies the $\TTT^{*}$-extension property, then $\fB\models\phi$.
\end{lem}
\begin{proof}
Fix a model $\fA$ of $\phi$.  
For each $k \in [0,\Width(\sigma)]$, let \(\TTT_k^{*} \eqdef \TTT^{\fA}_k\) be the set of $k$-types realised in~$\fA$.  
It is straightforward to verify that \(\TTT^{*} \;\eqdef\; \bigcup_{k=0}^{\Width(\sigma)} \TTT_k^{*}\) is closed and consistent.

Now suppose that $\fB$ is a $\sigma$-structure that is $\TTT^{*}$-guarded and satisfies the $\TTT^{*}$-extension property.  
We show that $\fB \models \phi$.

Let \(\forall \xs\,\big( \alpha_t(\xs) \rightarrow \exists \ys\,\big(\beta_t(\xs,\ys)\wedge \psi_t(\xs,\ys)\big)\big)\)
be a Skolem conjunct of~$\phi$, and let $f\colon \xs \to B$ be an assignment such that $\fB,f \models \alpha_t(\xs)$.  
Enumerate the elements of $f(\xs)\setminus \Cons$ as a tuple $\bs = \langle b_1,\dots,b_k \rangle$ without repetitions.
Because the atom $\alpha_t$ holds, the type $\type{\fB}{\bs}$ is guarded.
From the $\TTT^{*}$-guardedness of $\fB$, there exists a tuple $\as = \langle a_1,\dots,a_k \rangle$ of pairwise distinct unnamed elements of~$\fA$ such that \(\type{\fA}{\as} = \type{\fB}{\bs}.\)
Define an assignment $g\colon \xs \rightarrow A$ by
\[
  g(x) \;=\;
  \begin{cases}
    a_i &\text{if } f(x) = b_i,\\[2mm]
    f(x) &\text{if } f(x) \in \Cons.
  \end{cases}
\]

Since $\fA \models \phi$, and in particular $\fA,g \models \alpha_t(\xs)$, there exists an assignment  
$g' \colon \xs\ys \rightarrow A$ extending $g$ such that \(\fA,g' \models \beta_t(\xs,\ys) \wedge \psi_t(\xs,\ys).\)
Let $g'(\ys)\setminus (g(\xs) \cup \Cons) = \{a_{k+1},\dots,a_m\}$ be the unnamed elements introduced by~$g'$ (without repetitions), and form the extended tuple \(\as' = \langle a_1,\dots,a_m \rangle.\)

We next construct a tuple $\bs' = \langle b_1,\dots,b_m \rangle$ in~$\fB$ satisfying \(\type{\fB}{\bs'} = \type{\fA}{\as'}.\)
The first $k$ coordinates are already fixed, namely $b_i$ is as in $\bs$ for $i\le k$.  
For $i = k+1,\dots,m$, assume inductively that $b_1,\dots,b_{i-1}$ have been chosen.  
Since $\fB$ satisfies the $\TTT^{*}$-extension property and
\(\type{\fA}{a_1,\dots,a_{i-1}} = \type{\fB}{b_1,\dots,b_{i-1}},\)
there exists an element $b_i \in B \setminus \Cons$ distinct from $b_1,\dots,b_{i-1}$ such that
\(\type{\fB}{b_1,\dots,b_i}=\type{\fA}{a_1,\dots,a_i}.\)
Iterating this construction yields the required tuple $\bs'$.

Finally, define an assignment $f' \colon \xs\ys \rightarrow B$ extending~$f$ by setting
\[
  f'(y) \;=\;
  \begin{cases}
    b_i &\text{if } g'(y) = a_i,\\[2mm]
    g'(y) &\text{if } g'(y) \in \Cons.
  \end{cases}
\]
By construction of~$f'$, we obtain \(\fB,f' \models \beta_t(\xs,\ys) \wedge \psi_t(\xs,\ys).\)

Similar arguments establish the satisfaction of existential and universal conjuncts of~$\phi$.
\end{proof}

Let $\phi$ be a $\GF$-sentence in normal form.
If a family of sets of $k$-types $\TTT^* = \bigcup_{k=0}^{\Width(\sigma)}\TTT^*_k$ satisfies the conditions of Lemma~\ref{lemma:witness-of-satisfiability}, then we call it a \emph{satisfiability witness} for $\phi$.

W.l.o.g.~we will always assume $\TTT^*_k \neq \emptyset$ for all $k \in [0,\Width(\sigma)]$.
By Claim~\ref{claim:number-of-types}, we have that
\begin{flalign}
  |\TTT^*| = \sum_{k=0}^{\Width(\sigma)} |\TTT^{*}_k| \le \sum_{k=0}^{\Width(\sigma)} 2^{|\Rels| \cdot (k+|\Cons|)^{\Width(\sigma)}} = 2^{2^{\cO(|\phi|\cdot\log|\phi|)}},
\end{flalign}
provided that $\sigma = \Rels \uplus \Cons$ is the induced signature of $\phi$.

\subparagraph{Remarks on Computability.}
The proofs of Lemmas~\ref{lemma:GF-normal-form} and~\ref{lemma:witness-of-satisfiability} are inherently non-constructive, as their proofs work with an unspecified model of the sentence.
The same applies to the highlighted reduction to the standard name assumption (Section~\ref{sec:prelim}).
We briefly outline computational aspects of these reductions; in particular, required by Proposition~\ref{prop:constructive}.

Let $\phi$ be an arbitrary \GF{}-sentence over a signature $\sigma = \Rels \uplus \Cons$.
We first eliminate equalities between constants by considering all possible partitions $\mathcal{C}$ of the set~$\Cons$.
For each such partition we form a sentence $\phi_{\mathcal{C}}$ by identifying all constants within each block of~$\mathcal{C}$ and consistently replacing them by a fixed representative.

A similar idea allows us to make the normal-form reduction effective.
Rather than replacing subsentences by $\top$ or~$\bot$ according to their truth value in some unknown model,
we systematically consider all Boolean assignments $\mathcal{P}$ to the relevant subsentences of~$\phi$.
For each such assignment we obtain a normal-form sentence $\phi_{\mathcal{C},\mathcal{P}}$ from $\phi_{\mathcal{C}}$ in an expected way.

In this way we obtain an exponentially large family of normal-form \GF{}-sentences~$\phi_{\mathcal{C},\mathcal{P}}$.
The original sentence~$\phi$ is satisfiable iff at least one of these sentences is satisfiable under the standard name assumption.
Moreover, any model $\fA$ of some $\phi_{\mathcal{C},\mathcal{P}}$ can be transformed into a model $\fA'$ of~$\phi$ over the same domain by reinterpreting the constants in accordance with~$\mathcal{C}$ and then restricting to the original signature~$\sigma$.

Finally, for each sentence $\phi_{\mathcal{C},\mathcal{P}}$ one can decide in $2$-\ExpTime{} whether it is satisfiable under the standard name assumption and, if so, compute a corresponding satisfiability witness within the same complexity bound.
A particularly simple algorithm for this task is given in the book by Pratt-Hartmann~\cite{PH23};
although the definitions there differ slightly from ours, adapting the algorithm to our setting is straightforward.

\section{Probabilistic Model Constructions}
\label{sec:random}

In this section we establish Theorems~\ref{theorem:main} and~\ref{theorem:precise-bound}.  
We begin with an informal overview of the probabilistic approach (Subsection~\ref{sec:random-informal}),  
and then develop the technical details in two stages.  
First, in Subsection~\ref{sec:random-independent}, we present the baseline method of \emph{Independent Sampling},  
which already yields a doubly-exponential upper bound, sufficient for proving Theorem~\ref{theorem:main}.  
Next, in Subsection~\ref{sec:random-markov}, we refine the construction via \emph{Markovian Sampling},  
achieving a sharper bound that is necessary for proving Theorem~\ref{theorem:precise-bound}.  

\subsection{Informal Overview}
\label{sec:random-informal}

To gain intuition,
let us start with Gurevich and Shelah's probabilistic proof of the finite model property for the G\"odel Class~\cite{GS83}.
Consider a sentence $\phi$ in the shape of $\forall x_1,x_2\;\exists y\;\psi$, where $\psi$ is quantifier-free and uses only predicates of arity $1$ and $2$.
Assuming that $\phi$ has a model $\str{A}$, we generate a finite random structure $\fB$ as follows.
First prepare a domain $B$ of size $n \in \N$ and assign to its elements the $1$-types realised in $\str{A}$, so that each $1$-type is assigned roughly to the same number of elements.
Then we set the $2$-types: for each pair of distinct elements $b_1,b_2$ choose a $2$-type randomly from those $2$-types realised in $\str{A}$ whose endpoints agree with the already defined $1$-types of $b_1$ and $b_2$.
When $n$ is large enough, the structure $\fB$ becomes a finite model of $\phi$.
To see this, consider any two elements $b_1,b_2$ of $\fB$. Let $\langle a_1,a_2 \rangle$ be a pair in $\fA$ realising the same $2$-type as $\langle b_1,b_2 \rangle$.
With probability at least $1 - 2^{-\delta n}$, for a certain fixed independent of $n$ number $\delta>0$, some element $b_3$ of $\fB$ extends the $2$-type of $\langle b_1,b_2 \rangle$ to a $3$-type in the same way as a correct witness $a_3$ in $\fA$ extends the $2$-type of $\langle a_1,a_2 \rangle$.
By the union bound over all such pairs $\langle b_1,b_2 \rangle$, we get $\bbP[\fB \not\models \phi] \le n^2 \cdot 2^{-\delta n} \rightarrow 0$.

Imagine now a hypothetical generalisation of this method to sentences in the shape of $\phi = \forall x_1,x_2,x_3\;\exists y\;\psi$, where $\psi$ uses symbols of arbitrary arity.
We first assign $1$-types, then randomly choose $2$-types, \dots and when trying to assign $3$-types, we get stuck.
Indeed, the independent random choices could generate a configuration of $2$-types on pairs $\langle b_1,b_2 \rangle$, $\langle b_2,b_3 \rangle$, and $\langle b_1,b_3 \rangle$ which is not induced by any $3$-type realised in $\fA$.

Suppose now that $\phi$ is a guarded sentence in normal form.
If we decide not to put any facts speaking about $\{ b_1,b_2,b_3 \}$ or any of its supersets, then a guarded sentence cannot notice that this unintended pattern has occured.
Therefore, the following strategy appears natural. 
We first assign the $1$-types, then proceed by randomly choosing the $2$-types, and for the $3$-types we act as follows. 
For each triple $\langle b_1,b_2,b_3 \rangle$, we randomly select a $3$-type~$\tau$. 
However, before assigning $\tau$ to $\langle b_1,b_2,b_3 \rangle$, we first check whether $\tau$ is \emph{compatible} with the pattern induced by the previously defined $2$-types on $b_1,b_2,b_3$. (We make the notion of compatibility precise in a moment.)
If it is indeed compatible, then we set the $3$-type of $\langle b_1,b_2,b_3 \rangle$ to~$\tau$; 
otherwise, we simply omit this triple and continue with the remaining ones.
In a similar manner, we can then specify the $k$-types also for $k > 3$.

Assuming an appropriate choice of the parameter $n$ and relying on careful probability estimates,
we show that, although many tuples are omitted due to conflicts between $k$-types, there is still a significant chance that all required witnesses exist for all tuples.
Consequently, we conclude that there is a sequence of random choices that produces a finite model $\fB$ for $\phi$.

\subsection{Baseline: Independent Sampling}
\label{sec:random-independent}

In this subsection we prove Proposition~\ref{prop:weak-fmp-normal-form}.
Combined with Lemmas~\ref{lemma:GF-normal-form} and~\ref{lemma:witness-of-satisfiability}, it yields the upper bound of Theorem~\ref{theorem:main}: every satisfiable \GF{}-sentence $\phi$ admits a finite model whose domain has cardinality $2^{2^{\cO(|\phi|\cdot\log|\phi|)}}$.
(The lower bound of Theorem~\ref{theorem:main} is proven in Section~\ref{sec:tight}.)

\begin{prop}\label{prop:weak-fmp-normal-form}
  There exists a constant $C > 0$ such that the following holds.
  Let $\phi$ be a normal-form \GF{}-sentence with signature $\sigma = \Rels \uplus \Cons$.
  If~$\TTT^{*} \subseteq \TTT^{\sigma}$ is a satisfiability witness for $\phi$,
  then $\phi$ has a finite model with unnamed domain of size at most
  \[
    C \cdot |\TTT^{*}|^{2^{\Width(\sigma)}}.
  \]
\end{prop}

To prove Proposition~\ref{prop:weak-fmp-normal-form}, we describe a procedure that generates a $\sigma$-structure $\fB$ from a witness of satisfiability $\TTT^* = \bigcup_{k=0}^{\Width(\sigma)} \TTT^*_k$.  
The procedure is given in Algorithm~\ref{algos}.  
In addition to $\TTT^*$, the algorithm takes as input a parameter $n \in \N$, specifying the size of the unnamed part of the domain of $\fB$.  
The analysis of the algorithm is carried out in Lemma~\ref{lemma:algo-analysis}.

\begin{algorithm}[t]
  \caption{Randomised model generation from a satisfiability witness via \emph{Independent Sampling}}\label{algos}
  \KwIn{signature $\sigma = \Rels \uplus \Cons$,  
        satisfiability witness $\{\TTT^*_k\}_{k=0}^{\Width(\sigma)}$,  
        parameter $n \in \N$}
  \KwOut{random $\sigma$-structure $\fB$ with domain $B \eqdef \{1,2,\dots,n\} \uplus \Cons$}
  initialise $\fB$ as the empty $\sigma$-structure with domain $B$\;
  assign the unique $0$-type from $\TTT^*_0$ to $\fB$\;
  \For{$k = 1,2,\dots,\Width(\sigma)$\label{line:outer}}{
    \ForEach{$b_1,\dots,b_k \in [n]$ {\rm such that} $b_1 < \dots < b_k$\label{line:inner}}{
      choose a $k$-type $\tau \in \TTT^*_k$ uniformly at random\label{line:rand}\;
      \If{$\tau$ is compatible with $\type{\fB}{b_1,\dots,b_k}$\label{line:test}}{
        assign $\type{\fB}{b_1,\dots,b_k} \eqdef \tau$\label{line:modify}\;
      }
    }
  }
\end{algorithm}

In Algorithm~\ref{algos}, we rely on the following notion to determine a set of $k$-types to which a given configuration of atoms can be extended.
Let $k \in \N$ and let $\tau_1,\tau_2$ be $k$-types. 
We say that $\tau_1$ and $\tau_2$ are \emph{compatible} if these $k$-types agree on every literal that mentions a strict subset of the variables $\{x_1,\dots,x_k\}$, i.e., the interiors of $\tau_1$ and $\tau_2$ are the same: $\interior(\tau_1) = \interior(\tau_2)$.
For instance, $2$-types are compatible if they agree on the $1$-types corresponding to their endpoints $x_1$ and $x_2$ but possibly differ on some literals that use both $x_1$ and $x_2$; 
similarly $3$-types are compatible if they agree on the $2$-types induced on $\langle x_1,x_2 \rangle$, $\langle x_2,x_3 \rangle$, and $\langle x_1,x_3 \rangle$ but possibly differ on some literals that use simultaneously $x_1$, $x_2$, and $x_3$; 
and so on for larger $k$. 
In particular, $1$-types are compatible if they agree on constants.

\begin{lem}\label{lemma:algo-analysis}
  There exists a fixed constant $C > 0$ such that the following holds: if $\sigma$ is the signature of $\phi$,
  $\TTT^* = \bigcup_{k=0}^{\Width(\sigma)}\TTT^*_k$ is a satisfiability witness for $\phi$, and the parameter $n$ is chosen so that
  \[  n \ge C \cdot |\TTT^*|^{2^{\Width(\sigma)}}, \]
  then the structure $\fB$ generated by Algorithm~\ref{algos} satisfies $\fB \models \phi$ with probability at least $1/2$.
\end{lem}

\subparagraph{Proof of Lemma~\ref{lemma:algo-analysis}.}
W.l.o.g.~assume $|\TTT^*|\ge 3$.\footnote{If $\TTT^*=\{\tau_0\}$ with $\tau_0$ the unique $0$-type, then $\phi$ is satisfiable already in the domain $\Cons$. 
If $\TTT^*=\{\tau_0,\tau_1\}$ where $\tau_0$ is a $0$-type and $\tau_1$ a $1$-type, then $\phi$ is satisfiable in the domain $\Cons\cup\{1\}$.}
Set $\delta = |\TTT^*|^{-2^{\Width(\sigma)-1}}$.

For the analysis, we introduce a family of random variables
$\cX = \big\{ \cX(\bs) \mid \bs \in \bigcup_{k=1}^{\Width(\sigma)}\binom{[n]}{k} \big\}$.
For each $k \in [\Width(\sigma)]$ and each $k$-tuple $\bs \in \binom{[n]}{k}$,
the variable $\cX(\bs)$ records the $k$-type $\tau$ chosen by Algorithm~\ref{algos} at Line~\ref{line:rand} when processing $\bs$.
Thus, $\cX(\bs)$ is uniformly distributed over $\TTT^*_k$,
and the variables in $\cX$ are mutually independent.

Note that each $k$-tuple $\bs = \langle b_1,\dots,b_k \rangle \in \binom{[n]}{k}$ is assumed to be ordered increasingly.
For a permutation $\rho\colon [k] \rightarrow [k]$,
we will naturally write
\( \cX(b_{\rho(1)},\dots,b_{\rho(k)}) = \type{\cX(\bs)}{x_{\rho(1)},\dots,x_{\rho(k)}}, \)
i.e., $\cX(\rho(\bs))$ is the $k$-type $\cX(\bs)$ reindexed according to~$\rho$.  
This convention is adopted mainly for convenience in later proofs.
As the witness of satisfiability $\TTT^*$ is closed under permutations of $k$-types, this convention will not be problematic. 

\begin{clm}\label{claim:types-random}
  For every $k \in [\Width(\sigma)]$ and every $k$-tuple $\langle b_1, \dots, b_k \rangle \in \binom{[n]}{k}$, the following holds:
  \begin{enumerate}
    \item\label{types-random-item1} For any $k$-type $\tau \in \TTT_k^{*}$, we have $\type{\fB}{b_1,\dots,b_k} = \tau$ whenever
    \[
    \text{$\type{\tau}{x_{i_1}, \dots, x_{i_t}} = \cX(b_{i_1}, \dots, b_{i_t})$ for every $t \in [k]$ and every $1 \le i_1 < \dots < i_t \le k$.} \]
    \item\label{types-random-item2} If $\deltatype{\fB}{b_1,\dots,b_k} \neq \partial\tau_{\mathrm{all\text{-}neg}}$, then $\type{\fB}{b_1,\dots,b_k} = \cX(b_1,\dots,b_k)$.
  \end{enumerate}
\end{clm}
\begin{proof}
  Observe first that Algorithm~\ref{algos} initialises $\fB$ with facts only on the constants,
  and subsequently proceeds by monotonically adding new facts in each iteration.
  More precisely, when processing a tuple $\bs$, there are two possibilities:
  either $\fB$ remains unchanged, or the $k$-type on $\bs$ is set: $\type{\fB}{\bs} \eqdef \cX(\bs)$.
  Crucially, the latter occurs only if $\cX(\bs)$ is \emph{compatible} with the atoms already defined in $\fB$, i.e.,
  \(
    \interior(\type{\fB}{\bs}) = \interior(\cX(\bs)).
  \)
  Thus, only the boundary of $\type{\fB}{\bs}$---that is, the facts $R(\as)$ whose scope satisfies $\bs \subseteq \as \subseteq \bs \cup \Cons$---may be modified: $\deltatype{\fB}{\bs}$ is set to $\partial\cX(\bs)$.
  Moreover, boundaries corresponding to distinct tuples $\bs$ are disjoint, and therefore never interfere with each other.
  Consequently, for every $\bs$, we have either
  \[
    \deltatype{\fB}{\bs} = \partial\cX(\bs)
    \quad\text{and}\quad
    \interior(\type{\fB}{\bs}) = \interior(\cX(\bs)),
  \]
  or else $\deltatype{\fB}{\bs}$ remains equal to $\partial\tau_{\mathrm{all\text{-}neg}}$ throughout.

  The claim follows easily from the above explanation.
  Item~\ref{types-random-item1} can be proven by induction over the subtuples of $\bs = \langle b_1,\dots,b_k \rangle$.
  By the inductive hypothesis, every proper subtuple $\langle b_{i_1},\dots,b_{i_t} \rangle$ with $t \in [k - 1]$
  has already been assigned the $t$-type as prescribed by $\tau$, i.e., $\type{\fB}{b_{i_1},\dots,b_{i_t}} = \type{\tau}{x_{i_1},\dots,x_{i_t}}$.
  Since $\cX(\bs)=\tau$, the compatibility test at Line~\ref{line:test} succeeds, and the algorithm therefore assigns the type $\cX(\bs)$ to~$\bs$.
  Item~\ref{types-random-item2} is immediate.
\end{proof}

The main part of the proof of Lemma~\ref{lemma:algo-analysis} relies on three technical claims.  
Claim~\ref{claim:tau-guarded} establishes that $\fB$ is $\TTT^{*}$-guarded with certainty.  
Claims~\ref{claim:tau-extension-a} and~\ref{claim:tau-extension-b} establish that $\fB$ satisfies the $\TTT^{*}$-extension property with probability at least~$1/2$.
By Lemma~\ref{lemma:witness-of-satisfiability}, these two properties imply $\fB \models \phi$.

\begin{clm}\label{claim:tau-guarded}
  The structure $\fB$ is $\TTT^{*}$-guarded in every realisation, i.e., deterministically.
\end{clm}
\begin{proof}
  It is readily verified that the $0$-type of $\fB$ is the unique element of $\TTT^{*}_0$.
  Likewise, every $1$-type realised in $\str{B}$ comes from $\TTT^{*}_1$: for every $b \in [n]$, its $1$-type $\type{\str{B}}{b}$ is given by $\cX(b) \in \TTT^{*}_1$.

  For $k \in [2,\Width(\sigma)]$, we fix a $k$-tuple $\bs  \in \binom{[n]}{k}$ such that its $k$-type $\type{\fB}{\bs}$ is guarded, i.e., it contains a positive literal $\gamma$ with $\FreeVars(\gamma) = \{ x_1,\dots,x_k \}$.
  As $\gamma$ belongs to the boundary of $\type{\fB}{\bs}$, we have that $\deltatype{\fB}{\bs} \neq \partial\tau_{\mathrm{ all\text{-}neg}}$.
  From Claim~\ref{claim:types-random}, it follows that $\type{\fB}{\bs} = \cX(\bs) \in \TTT^{*}_k$.
\end{proof}

\begin{clm}\label{claim:tau-extension-a}
  Let $k \in [0,\Width(\sigma){-}1]$ and let $\bs \in \binom{[n]}{k}$.
  Fix a $k$-type $\tau_1 \in \TTT^{*}_k$ and a $(k{+}1)$-type $\tau_2 \in \TTT^{*}_{k+1}$ satisfying that $\tau_2 \models \tau_1$.
  Define $\Fail(\tau_2 \models \tau_1; \bs)$ to be the probabilistic event that
  \[
    \type{\fB}{\bs} = \tau_1
    \quad \text{and} \quad
    \type{\fB}{\bs,b'} \neq \tau_2 \text{ for every } b' \in \textstyle [n] \setminus \bs.
  \]
  Then \(\bbP[ \Fail(\tau_2 \models \tau_1; \bs) ] \le e^{- \delta \cdot (n - \Width(\sigma))}\).
\end{clm}
\begin{proof}
  Assume that $\bbP[ \type{\fB}{\bs} = \tau_1 ] > 0$, as otherwise $\bbP[ \Fail(\tau_2 \models \tau_1; \bs) ] = 0$ and the claim holds trivially. We condition on the probabilistic event $\{ \type{\fB}{\bs} = \tau_1 \}$.

  Enumerate the elements of $\bs$ as $b_1,\dots,b_k$,
  and fix $b_{k+1} \in [n] \setminus \bs$. We argue that, with probability at least $\delta$, it happens that $\{\type{\fB}{\bs,b_{k+1}} = \tau_2 \}$.
  From Claim~\ref{claim:types-random},
  it follows that $\type{\fB}{b_1,\dots,b_{k+1}} = \tau_2$ whenever,
  for every $t \in [k{+}1]$ and every $1 \le i_1 < \dots < i_t \le k + 1$,
  we have that $\cX(b_{i_1},\dots,b_{i_t}) = \type{\tau_2}{x_{i_1},\dots,x_{i_t}}$.
  Since $\tau_2 \models \tau_1$ and $\{ \type{\fB}{\bs} = \tau_1 \}$ is fixed,
  we can restrict our attention to subsequences having $i_t = k+1$.
  Consequently, the random event $\{ \type{\fB}{\bs,b_{k+1}} = \tau_2 \}$, under the condition $\{ \type{\fB}{\bs} = \tau_1 \}$, can be expanded as
  \begin{align}\label{eqn:expanded-event}
    \bigcap_{\substack{0 \le t \le k \\ 1 \le i_1 < \dots < i_t \le k}} \big\{ \, \cX(b_{i_1},\dots,b_{i_t},b_{k+1}) = \type{\tau_2}{x_{i_1},\dots,x_{i_t},x_{k+1}} \, \big\}.
  \end{align}
  Using the independence of variables in $\cX$, the probability $\bbP[ \type{\fB}{\bs,b_{k+1}} = \tau_2 \mid \type{\fB}{\bs} = \tau_1 ]$ can be lower bounded by $\delta$ as follows:
  \begin{align}\label{eqn:iloczyn-pstwo}
    &\prod_{\substack{0 \le t \le k \\ 1 \le i_1 < \dots < i_t \le k}}
    \bbP\big[ \cX(b_{i_1},\dots,b_{i_t},b_{k+1}) = \type{\tau_2}{x_{i_1},\dots,x_{i_t},x_{k+1}} \bigmid \type{\fB}{\bs} = \tau_1 \big] \nonumber \\
    &= \prod_{t=0}^{k} |\TTT^*_{t+1}|^{-\binom{k}{t}} \ge \prod_{t=0}^{k} |\TTT^*|^{-\binom{k}{t}} = |\TTT^*|^{-2^k} \ge |\TTT^*|^{-2^{\Width(\sigma)-1}} = \delta. 
  \end{align}

  \smallskip
  We keep the $k$-tuple $\bs$, the $k$-type $\tau_1$, and the $(k{+}1)$-type $\tau_2$ fixed as before,
  as well as the condition $\{ \type{\fB}{\bs} = \tau_ 1\}$.
  We now bound the probability that no candidate $b'$ for $b_{k+1}$ makes $\type{\fB}{\bs,b_{k+1}} = \tau_2$.
  Since $\{ \type{\fB}{\bs} = \tau_ 1\}$ is fixed, the events
  $\{ \type{\fB}{\bs,b'} = \tau_2 \}$, where $b'$ ranges the set $[n] \setminus \bs$, are mutually independent.
  Indeed, by~\eqref{eqn:expanded-event}, they are generated by disjoint subsets of variables from $\cX$.
  Hence, the event $\Fail(\tau_2 \models \tau_1; \bs)$ happens with probability at most
  \begin{equation}\label{eqn:b}
    \prod_{b' \in [n] \setminus \{b_1,\dots,b_k\}} \bbP\big[ \type{\fB}{\bs,b'} \neq \tau_2 \bigmid \type{\fB}{\bs} = \tau_1 \big] \le (1 - \delta)^{n - k} \le e^{- \delta \cdot (n - \Width(\sigma))}.
  \end{equation}
  In the last inequality, we use that $1 - \delta \le e^{-\delta}$, for any $\delta \in \R$, to move $\delta$ to the exponent.

  We conclude that $\bbP[ \Fail(\tau_2 \models \tau_1; \bs) ] \le e^{- \delta \cdot (n - \Width(\sigma))}$, as claimed.
\end{proof}

\begin{clm}\label{claim:tau-extension-b}
  Let $K = 8 \cdot \delta^{-1} \cdot (\Width(\sigma) + \ln|\TTT^*|)$. 
  If \( n \ge K \cdot \ln K, \) then the structure~$\fB$ satisfies the $\TTT^{*}$-extension property with probability at least \( 1/2 \). 
\end{clm}
\begin{proof}
  The structure~$\fB$ satisfies the $\TTT^{*}$-extension property precisely when none of the events $\Fail(\tau_2 \models \tau_1; \bs)$ occur.
  Applying the union bound, we estimate the probability of the complementary event—that at least one of the events $\Fail(\tau_2 \models \tau_1; \bs)$ occurs—as follows:
  \begin{equation}\label{eqn:cc}
    \sum_{\langle \bs, \tau_1, \tau_2 \rangle \in \Omega}
    \bbP\big[\Fail(\tau_2 \models \tau_1; \bs)\big]
    \le
    n^{\Width(\sigma)} \cdot |\TTT^{*}| \cdot e^{-\delta \cdot (n - \Width(\sigma))},
  \end{equation}
  where the summation ranges over the set
  \begin{equation}
    \Omega =
    \bigcup_{k=0}^{\Width(\sigma)-1}
    \big\{
      \langle \bs, \tau_1, \tau_2 \rangle
      \in \textstyle{\binom{[n]}{k}} \times \TTT^{*}_{k} \times \TTT^{*}_{k+1}
      \bigmid
      \tau_2 \models \tau_1
    \big\}.
  \end{equation}

  The notation $\bs \in \binom{[n]}{k}$ implicitly assumes that $\bs$ is sorted in increasing order.
  Since $\TTT^*$ is closed, it suffices to consider the events $\Fail(\tau_2 \models \tau_1; \bs)$ only for such tuples.

  By Claim~\ref{claim:tau-extension-a}, we have the estimate on probability: \(\bbP[\Fail(\tau_2 \models \tau_1; \bs)] \le e^{-\delta \cdot (n - \Width(\sigma))}.\)
  Since $\tau_2$ determines $\tau_1$ and \(\sum_{k=0}^{\Width(\sigma)-1} \binom{n}{k} \le n^{\Width(\sigma)}\) for all \(n \ge 2\), we have the estimate on size: $|\Omega| \le n^{\Width(\sigma)} \cdot |\TTT^{*}|$.
  Combining these estimates yields~\eqref{eqn:cc}.

  Let $p$ denote the right side of~\eqref{eqn:cc}.
  We require $p \le 1/e < 1/2$.
  Taking natural logarithms on both sides of $p \le 1/e$ yields
  \begin{equation}\label{eqn:dd}
    \Width(\sigma)\cdot\ln n + \ln|\TTT^*| - \delta\cdot\big(n-\Width(\sigma)\big) \le -1.
  \end{equation}
  Rearranging~\eqref{eqn:dd} to isolate $n$ gives
  \begin{equation}\label{eqn:ee}
    n \ge \delta^{-1}\cdot\big(\Width(\sigma)\cdot\ln n + \ln|\TTT^*| + 1 \big) + \Width(\sigma).
  \end{equation}
  By collecting the terms in~\eqref{eqn:ee} into two groups $\nu$ and $\mu$, 
  we can rewrite the inequality as
  \begin{equation*}\label{eqn:nu-oraz-mu}
    n \ge \nu \cdot \ln n + \mu,
    \quad \text{where} \quad
    \nu = \delta^{-1}\cdot\Width(\sigma)
    \quad \text{and} \quad
    \mu = \delta^{-1}\cdot\big(\ln|\TTT^*| + 1\big) + \Width(\sigma).
  \end{equation*}
  The inequality $n \ge \nu \cdot \ln n + \mu$ holds whenever 
  \(
    n \ge 4\cdot(\nu + \mu)\cdot\ln(\nu + \mu).
  \)
  Moreover,
  \[
    \nu + \mu
    = \delta^{-1}\cdot(\Width(\sigma) + \ln|\TTT^*| + 1) + \Width(\sigma)
    \le 2 \cdot \delta^{-1} \cdot (\Width(\sigma) + \ln|\TTT^*|) \le K/4.
  \]
  It follows that choosing $n \ge K \cdot \ln K$ ensures $p \le 1/e$,
  thereby establishing the claim.
\end{proof}

We conclude now the proof of Lemma~\ref{lemma:algo-analysis}.
Since $|\TTT^*| \ge 3$ and $\delta^{-1} = |\TTT^*|^{2^{\Width(\sigma)-1}}$, it holds $\Width(\sigma) \le \ln(\delta^{-1})$ and $\ln|\TTT^*| \le \ln(\delta^{-1})$.
In consequence, the number $K$ from Claim~\ref{claim:tau-extension-b} satisfies $K \le 16 \cdot \delta^{-1}\cdot \ln \delta^{-1}$.
Hence there exists a constant $C>0$ such that
\begin{equation}\label{eqn:almost}
  K \cdot \ln K \le 16 \cdot\delta^{-1}\cdot \underbrace{\ln \delta^{-1} \cdot \ln\big(16 \cdot\delta^{-1}\cdot\ln \delta^{-1} \big)}_{\text{asymptotically dominated by } \delta^{-1}} \le C \cdot \delta^{-2}.
\end{equation}
We derive the threshold of Lemma~\ref{lemma:algo-analysis} directly from Claim~\ref{claim:tau-extension-b} and inequality~\eqref{eqn:almost}:
\begin{equation}
  n \ge C \cdot \delta^{-2} = C \cdot |\TTT^*|^{2 \cdot 2^{\Width(\sigma)-1}} = C \cdot |\TTT^*|^{2^{\Width(\sigma)}}.
\end{equation}

\subparagraph{Additional Remarks.}
We conclude this subsection with some further remarks.

\begin{rem}
  The converse of Item~\ref{types-random-item1} of Claim~\ref{claim:types-random} does not necessarily hold.
  For example, let $\tau \in \TTT^{*}_3$ be a $3$-type such that
  $\deltatype{\tau}{x_1,x_2} = \partial\tau_{\mathrm{all\text{-}neg}}$.
  It may still happen that Algorithm~\ref{algos} assigns $\tau$ to some triple $\langle a_1,a_2,a_3 \rangle$,
  even though the condition of Item~\ref{types-random-item1} is not satisfied.

  Indeed, suppose that the $1$-types of $a_1$, $a_2$, and $a_3$ are assigned according to $\tau$,
  and likewise the $2$-types for $\langle a_2,a_3 \rangle$ and $\langle a_1,a_3 \rangle$.
  However, let $\cX(a_1,a_2)$ be a $2$-type distinct from
  $\type{\tau}{x_1,x_2}$,
  and assume that the corresponding $1$-types are incompatible with those already defined on $a_1$ and $a_2$,
  say $\type{\cX(a_1,a_2)}{x_1} \neq \type{\tau}{x_1}$.
  In this case, Algorithm~\ref{algos} rejects $\cX(a_1,a_2)$
  (the test at Line~\ref{line:test} fails).
  Nevertheless, in the subsequent iteration,
  the $3$-type $\tau$ may still be selected for $\langle a_1,a_2,a_3 \rangle$, as $\tau$ and $\type{\str{B}}{a_1,a_2,a_3}$ remain compatible.

  This subtle behaviour, however, does not affect the correctness of our later arguments.
\end{rem}

\begin{rem}
  One may wonder whether our choice of $\delta = |\TTT^*|^{-2^{\Width(\sigma)-1}}$ is overly pessimistic,
  and if a larger value of $\delta$ could yield a sharper analysis.
  The crucial role of $\delta$ is to provide a lower bound on the product in~\eqref{eqn:iloczyn-pstwo}.
  However, under a reasonable assumption $|\TTT^*_k| \approx |\TTT^{\sigma}_k|$, we show that an exponential power of $|\TTT^*|$ is in fact unavoidable.\footnote{The formulation of Proposition~\ref{prop:weak-fmp-normal-form} does not constrain the length of the sentence~$\phi$.
  Consequently, it is easy to construct sentences that enforce the realisation of every guarded $k$-type. 
  For a more refined perspective, see Corollary~\ref{corollary:precise-tight}, which shows that even sentences of length~$\cO(k)$ can require the realisation of nearly all $k$-types—albeit only at the doubly logarithmic scale, i.e., $\log_2\log_2 |\TTT_k^*| \ge (1-\varepsilon) \cdot \log_2\log_2|\TTT_k^{\sigma}|$.}
\end{rem}

\begin{clm}\label{eqn:better-delta}
  For every $w \ge 5$,
  there exists a signature $\sigma$ of width $w$ such that
  \[
    \prod_{i=0}^{w-1} |\TTT^{\sigma}_{i+1}|^{\binom{w-1}{i}} \ge |\TTT^{\sigma}_w|^{\,2^{w / 10} \cdot (w+1)^{-2}}.
  \]
\end{clm}
\begin{proof}
  Fix $w \ge 5$.
  Consider a signature $\sigma = \{ R \}$ such that the relation symbol $R$ has arity~$w$.
  From Claim~\ref{claim:number-of-types}, it follows that $|\TTT^{\sigma}_k| = 2^{k^w}$ for all $k \in \N$. 

  For every index $i \in [w-1]$, set $x_i = (i+1)/w$.
  We have the following estimation~\cite{cover2006elements}:
  \begin{equation}
    \binom{w-1}{i} = \frac{i + 1}{w} \cdot \binom{w}{i + 1} \ge \frac{i + 1}{w \cdot (w+1)} \cdot 2^{w \cdot H(x_i)},
  \end{equation}
  where $H(x) = - x\cdot \log_2(x) - (1-x)\cdot\log_2(1-x)$ is the binary entropy function.

  We lower bound the logarithm of the $i$th term of the product $\prod_{i=0}^{w-1} |\TTT^{\sigma}_{i+1}|^{\binom{w-1}{i}}$:
  \begin{equation}
    \log_2\Big(|\TTT^{\sigma}_{i+1}|^{\binom{w-1}{i}}\Big) = \binom{w-1}{i} \cdot (i+1)^w
    \ge \frac{i + 1}{w \cdot (w+1)} \times (i + 1)^w \cdot 2^{w \cdot H(x_i)}.
  \end{equation}
  We rewrite the last term $(i + 1)^w \cdot 2^{w \cdot H(x_i)}$ as follows:
  \begin{equation}
  (x_i \cdot w)^w \cdot 2^{w \cdot H(x_i)}
  = 2^{w \cdot H(x_i) + w\cdot\log_2 x_i} \cdot w^w
  = 2^{w \cdot (1-x_i) \cdot\log_2(\tfrac{x_i}{1-x_i})} \cdot w^w.
  \end{equation}
  Define $h(x) = (1-x) \cdot\log_2(\tfrac{x}{1-x})$.
  One can verify that $h(x) \ge 1/10$ for all $x \in (6/10,9/10)$.
  Assuming $w \ge 5$, we can always choose $j \in [w-1]$ satisfying $x_{j} \in (6/10,9/10)$.
  Thus
  \begin{equation}
    \log_2\Big(|\TTT^{\sigma}_{j+1}|^{\binom{w-1}{j}}\Big) 
    \ge \frac{j + 1}{w \cdot (w+1)} \times 2^{w/10} \cdot w^w
    \ge \log_2\Big(|\TTT^{\sigma}_w|^{2^{w/10} \cdot (w+1)^{-2}}\Big).
  \end{equation}
  Finally, since each term of the product $\prod_{i=0}^{w-1} |\TTT^{\sigma}_{i+1}|^{\binom{w-1}{i}}$ is at least $1$,
  we conclude that
  \begin{equation}
    \prod_{i=0}^{w-1} |\TTT^{\sigma}_{i+1}|^{\binom{w-1}{i}} \ge |\TTT^{\sigma}_{j+1}|^{\binom{w-1}{j}} \ge |\TTT^{\sigma}_w|^{2^{w/10} \cdot (w+1)^{-2}},
  \end{equation}
  as claimed.
\end{proof}

\begin{rem}
  Gr\"adel~\cite{Gra99} showed that \GF{} admits lower complexity when either the predicate arity or the number of variables is bounded. More precisely, under these restrictions, the satisfiability problem for \GF{} is \ExpTime-complete instead of $2$-\ExpTime-complete.
  We observe that Proposition~\ref{prop:weak-fmp-normal-form} yields correspondingly smaller models in this setting as well.

  If the predicate arity is bounded (i.e., the width of the signature is bounded by a constant), then,
  by Claim~\ref{claim:number-of-types}, the size of witnesses becomes
  $|\TTT^*| \le 2^{p(|\phi|)}$, where $p$ is a fixed polynomial of degree $\Width(\sigma)$.
  Consequently, the bound from Proposition~\ref{prop:weak-fmp-normal-form},
  \(|\TTT^*|^{2^{\Width(\sigma)}},\)
  is singly exponential in the length of the sentence, as $2^{\Width(\sigma)}$ is at most a constant.

  The finite-variable case requires a more subtle argument.
  Indeed, in the presence of a predicate $R$ of arity $n$ and a constant symbol $c$, the number of $k$-types is already doubly exponential for $k = 1$.
  This is because a $1$-type must contain a literal for every atom of the form $R(\vs)$, where $\vs \in \{x_1, c\}^n$.

  To obtain a singly exponential upper bound, we restrict attention to \emph{$\phi$-active} atoms.
  An atom $R(\as)$ is called $\phi$-active if $\phi$ contains an atom $R(\bs)$ such that $\as = f(\bs)$ for some substitution $f \colon \bs \to \{ x_1,\dots,x_k \} \cup \Cons$ being the identity on constants, $f(c) = c$.
  Since $\phi$ cannot quantify over atoms that are not $\phi$-active, w.l.o.g., every $k$-type occurring in a satisfiability witness contains negative literals for non-active atoms.
  It follows that the number of $\phi$-active $k$-types is bounded by
  \(
    2^{|\phi| \cdot (k + |\Cons|)^{|V|}},
  \)
  where $V$ denotes the set of variables occurring in $\phi$.
  Since $|V|$ is bounded, the size of witnesses, $|\TTT^*|$, is singly exponential in~$|\phi|$.
  Finally,
  in the model-size bound from Proposition~\ref{prop:weak-fmp-normal-form}, \(|\TTT^*|^{2^{\Width(\sigma)}}\),
  the exponent $2^{\Width(\sigma)}$ is replaced by $2^{|V|}$:
  there is no need to consider tuples with more than $|V|$ unnamed elements.
\end{rem}

\subsection{Alternative: Markovian Sampling}
\label{sec:random-markov}

Proposition~\ref{prop:weak-fmp-normal-form} yields Theorem~\ref{theorem:main}, but
is not strong enough to yield
Theorem~\ref{theorem:precise-bound}.
Indeed, Theorem~\ref{theorem:precise-bound} demands an upper bound on the domain size that grows like
$|\TTT^{\sigma}_{\Width(\sigma)}|^{\,1-1/e+\varepsilon}$,
with $\varepsilon \to 0$ as the width $\Width(\sigma)$ tends to infinity.
In contrast, Claim~\ref{eqn:better-delta} shows that Algorithm~\ref{algos} developed in Subsection~\ref{sec:random-independent} may, in the worst case, produce models whose size grows much faster, namely as
$|\TTT^{\sigma}_{\Width(\sigma)}|^{\,2^{\Omega(\Width(\sigma))}}$.
Thus, a more delicate argument is required in order to obtain the sharper bound
stated in Theorem~\ref{theorem:precise-bound}.
In this subsection, we therefore refine Algorithm~\ref{algos} and establish
Proposition~\ref{prop:strong-fmp-normal-form}.
Theorem~\ref{theorem:precise-bound} then follows by combining it with
Lemma~\ref{lemma:ratio-types}, stated and proved subsequently.

\begin{prop}\label{prop:strong-fmp-normal-form}
  There exists a constant $C > 0$ such that the following holds.
  Let $\phi$ be a normal-form \GF{}-sentence with signature $\sigma = \Rels \uplus \Cons$.
  If~$\phi$~is satisfiable, then it has a finite model with unnamed domain of size at most
  \[
    C \cdot \frac{|\TTT_{\Width(\sigma)}^{\sigma}|}{|\TTT_{\Width(\sigma)-1}^{\sigma}|} \cdot \big(\ln |\TTT_{\Width(\sigma)}^{\sigma}|\big)^2.
  \]
\end{prop}
  
Let us first see where the bottleneck of Algorithm~\ref{algos} occurs.
When processing a tuple $\bs$, Algorithm~\ref{algos} samples a $k$-type $\tau$ uniformly at random, independently of all previous choices.  
Only a posteriori it verifies whether the chosen $\tau$ is compatible with the already defined structure on $\type{\fB}{\bs}$.  
This leads to a loss of probability, as many samples are rejected.

To improve this, we incorporate information from previous decisions directly into the sampling process.  
The key idea is to replace independent sampling with a Markovian scheme: when considering a tuple $\bs$, we first determine whether the pattern induced by the already fixed structure on $\type{\fB}{\bs}$ can be extended to a $k$-type in~$\TTT_k^*$.  
If so, we choose uniformly at random among the $k$-types that are compatible with this pattern; if not, we simply omit the tuple $\bs$, forcing all atoms involving precisely those unnamed elements to be false.  
In this way, every random choice has a higher chance of being accepted, boosting the overall success probability of the procedure.  

We formalise this idea in Algorithm~\ref{algos2}, and carry out the analysis in Lemma~\ref{lemma:algo2-analysis}.

\begin{algorithm}[t]
  \caption{Randomised model generation from a satisfiability witness via \emph{Markovian Sampling}}\label{algos2}
  \KwIn{signature $\sigma = \Rels \uplus \Cons$,  
        satisfiability witness $\{\TTT^*_k\}_{k=0}^{\Width(\sigma)}$,  
        parameter $n \in \N$}
  \KwOut{random $\sigma$-structure $\fB$ with domain $B \eqdef \{1,2,\dots,n\} \uplus \Cons$}
  initialise $\fB$ as the empty $\sigma$-structure with domain $B$\;
  assign the unique $0$-type from $\TTT^*_0$ to $\fB$\label{line2:zero}\;
  \For{$k = 1,2,\dots,\Width(\sigma)$\label{line2:outer}}{
    \ForEach{$b_1,\dots,b_k \in [n]$ {\rm such that} $b_1 < \dots < b_k$\label{line2:inner}}{
      let $\TTT$ be the set $\{ \, \tau \in \TTT^*_k \mid \tau \text{\emph{ is compatible with }} \type{\fB}{b_1,\dots,b_k} \, \}$\label{line2:set}\;
      \If{$\TTT \neq \emptyset$\label{line2:test}}{
        choose a $k$-type $\tau \in \TTT$ uniformly at random\label{line2:rand}\;  
        assign $\type{\fB}{b_1,\dots,b_k} \eqdef \tau$\label{line2:modify}\;
      }
    }
  }
\end{algorithm}

\begin{lem}\label{lemma:algo2-analysis}
  There exists a fixed constant $C > 0$ such that the following holds:
  if $\sigma$ is the signature of $\phi$, $\TTT^* = \bigcup_{k=0}^{\Width(\sigma)}\TTT^*_k$ is a satisfiability witness for $\phi$, and the parameter $n$ is chosen so that
  \[  n \ge C \cdot \frac{|\TTT^{\sigma}_{\Width(\sigma)}|}{|\TTT^{\sigma}_{\Width(\sigma)-1}|} \cdot \big(\ln |\TTT^{\sigma}_{\Width(\sigma)}|\big)^2, \]
  then the structure $\fB$ generated by Algorithm~\ref{algos2} satisfies $\fB \models \phi$ with probability at least $1/2$.
\end{lem}

\subparagraph{Proof of Lemma~\ref{lemma:algo2-analysis}.}
W.l.o.g.~$|\TTT^*| \ge 3$.
Set
\[ \delta_1 = \prod_{k=1}^{\Width(\sigma)} \big|\partial\TTT^{*}_{k}\big|^{-\binom{\Width(\sigma)-1}{k - 1}}. \]

For the analysis, we again define the random variables
\(\cX = \big\{ \cX(\bs) \;\big|\; \bs \in \bigcup_{k=1}^{\Width(\sigma)} \binom{[n]}{k} \big\}.\)
For each $k \in [\Width(\sigma)]$ and each $k$-tuple $\bs \in \binom{[n]}{k}$,
the variable $\cX(\bs)$ records the $k$-type $\tau$ chosen by Algorithm~\ref{algos2} at Line~\ref{line2:rand} when processing~$\bs$.
If the non-emptiness test at Line~\ref{line2:test} fails, we set $\cX(\bs) \eqdef \tau_{\mathrm{all\text{-}neg}}$, the unique $k$-type consisting only of negative literals.
We again complete the set $\cX$ under permutations, i.e., write \(\cX(\rho(\bs)) = \type{\cX(\bs)}{x_{\rho(1)},\dots,x_{\rho(k)}} \)
for a permutation $\rho$.

Unlike in Algorithm~\ref{algos}, the random variables $\cX$ are no longer mutually independent: the set of admissible $k$-types at Line~\ref{line2:set} depends on previous random choices.
More precisely, for every $k$-tuple $\bs \in \binom{[n]}{k}$, the variable $\cX(\bs)$ is a random function of $\{ \cX(\as) \mid \as \in \bigcup_{i=1}^{k-1}\binom{\bs}{i} \}$.

The proof of Lemma~\ref{lemma:algo2-analysis} follows the same general scheme as that of Lemma~\ref{lemma:algo-analysis}.
In particular, Claims~\ref{claim:types-random} and~\ref{claim:tau-guarded} remain valid without modification.
We now turn to the analogue of Claim~\ref{claim:tau-extension-a}, denoted Claim~\ref{claim:tau-extension-markov}, whose proof requires exploiting the Markov property in place of full independence.

\begin{clm}\label{claim:tau-extension-markov}
  Let $k \in [0,\Width(\sigma){-}1]$, and let $\bs = \langle b_1,\dots,b_k \rangle \in \binom{[n]}{k}$.
  Fix a $k$-type $\tau_1 \in \TTT^{*}_k$ and a $(k{+}1)$-type $\tau_2 \in \TTT^{*}_{k+1}$ satisfying that $\tau_2 \models \tau_1$.
  Then \(\bbP[ \Fail(\tau_2 \models \tau_1; \bs) ] \le e^{- \delta_1 \cdot (n - \Width(\sigma))}\).
\end{clm}
\begin{proof}
  Assume that $\bbP[ \type{\fB}{\bs} = \tau_1 ] > 0$. We condition on the random event $\{ \type{\fB}{\bs} = \tau_1 \}$.

  Let us fix an element $b_{k+1} \in [n] \setminus \{ b_1,\dots,b_k \}$.
  We show that $\{ \type{\fB}{\bs,b_{k+1}}=\tau_2 \}$ happens with probability at least~$\delta_1$.

  Let $I_1,\dots,I_{2^k-1}$ enumerate the non-empty subsets of $[k]$ in non-decreasing order of size.
  Define $I_{2^k} \eqdef \{ k+1 \}$. For $r \in [2^k+1,2^{k+1}-1]$, set $I_{r} \eqdef I_{r - 2^k} \cup \{ k+1 \}$.

  For each set $I_r$, denoting $t = |I_r|$ and assuming that $i_1<\dots<i_t$ are the elements of $I_r$, we define the random event
  \(\cE(I_r) \eqdef \big\{ \type{\fB}{b_{i_1},\dots,b_{i_t}} = \type{\tau_2}{x_{i_1},\dots,x_{i_t}} \big\} \).
  We claim
  \begin{equation}\label{eqn:markov-prob}
    \bbP\big[ \cE(I_r) \bigmid \bigcap_{s < r} \cE(I_s) \big]
    \ge \big| \partial\TTT^{*}_t \big|^{-1}.
  \end{equation}

  From the condition $\bigcap_{s < r} \cE(I_s)$, it follows
  that every proper subtuple of $\langle b_{i_1},\dots,b_{i_t} \rangle$ has already assigned the type as prescribed by $\tau_2$.
  Thus, the $t$-type $\type{\tau_2}{x_{i_1},\dots,x_{i_t}}$ is compatible with the current type $\type{\fB}{b_{i_1},\dots,b_{i_t}}$, i.e., their interiors coincide,
  and it is available in the set of $t$-types defined at Line~\ref{line2:set}.  
  The check at Line~\ref{line2:test} therefore succeeds, and Algorithm~\ref{algos2} selects this $t$-type with probability inversely proportional to the size of this set:
  \begin{align}
    \bbP\big[ \cE(I_r) \bigmid \bigcap_{s < r} \cE(I_s) \big] &= \bbP\big[ \cX(b_{i_1},\dots,b_{i_t}) = \type{\tau_2}{x_{i_1},\dots,x_{i_t}} \bigmid \bigcap_{s < r} \cE(I_s) \big] \nonumber\\
    &= \big|\big\{ \tau \in \TTT^{*}_t \bigmid \interior(\tau) = \inttype{\tau_2}{x_{i_1},\dots,x_{i_t}} \big\}\big|^{-1}
    \ge \big| \partial\TTT^{*}_t \big|^{-1}.
  \end{align}

  Having~\eqref{eqn:markov-prob} established, we use the chain rule for probabilities and obtain
  \begin{equation}
    \bbP\big[\bigcap_{i=1}^{2^{k+1}-1} \cE(I_i) \bigmid \bigcap_{j=1}^{2^k-1} \cE(I_j) \big] =
    \prod_{i=2^k}^{2^{k+1}-1} \bbP\big[\cE(I_i) \bigmid \bigcap_{j<i} \cE(I_j) \big] \ge
    \prod_{t=1}^{k+1} \big|\partial\TTT^{*}_{t}\big|^{-\binom{k}{t-1}} \ge \delta_1.
  \end{equation}
  In the last inequality, we use $\big|\partial\TTT^{*}_{t}\big|^{-\binom{k}{t-1}} \ge \big|\partial\TTT^{*}_{t}\big|^{-\binom{\Width(\sigma)-1}{t-1}}$ to bound the product by $\delta_1$.

  We conclude $\bbP[ \type{\fB}{\bs,b_{k+1}} = \tau_2 \mid \type{\fB}{\bs} = \tau_1 ] \ge \delta_1$, as it holds that \[\bigcap_{i=1}^{2^{k+1}-1} \cE(I_i) = \{ \type{\fB}{\bs,b_{k+1}} = \tau_2 \},
  \; \text{and since $\tau_2 \models \tau_1$, also} \; \bigcap_{j=1}^{2^k-1} \cE(I_j) = \{ \type{\fB}{\bs} = \tau_1 \}. \]

  \smallskip
  Keeping the $k$-tuple $\bs$, the $k$-type $\tau_1$, and the $(k{+}1)$-type $\tau_2$ fixed,
  as well as the condition $\{ \type{\fB}{\bs} = \tau_ 1\}$,
  we bound the probability that no candidate $b'$ for $b_{k+1}$ makes $\type{\fB}{\bs,b_{k+1}} = \tau_2$.
  The events
  $\{ \type{\fB}{\bs,b'} = \tau_2 \}$, where $b'$ ranges the set $[n] \setminus \bs$, are mutually independent,
  as their common part $\{ \type{\fB}{\bs} = \tau_1 \}$ is fixed.
  We thus have
  \begin{equation}
    \prod_{b' \in [n] \setminus \{b_1,\dots,b_k\}} \bbP\big[ \type{\fB}{\bs,b'} \neq \tau_2 \bigmid \type{\fB}{\bs} = \tau_1 \big] \le (1 - \delta_1)^{n - k} \le e^{- \delta_1 \cdot (n - \Width(\sigma))},
  \end{equation}
  concluding the proof of the claim.
\end{proof}

Having established Claim~\ref{claim:tau-extension-markov},  
we may now invoke Claim~\ref{claim:tau-extension-b} with $\delta \eqdef \delta_1$,  
since its proof requires only that $e^{- \delta_1 \cdot (n - \Width(\sigma))}$ bounds the probability of the event  
$\Fail(\tau_2 \models \tau_1; \bs)$ and $|\TTT^*| \ge 3$.  
Hence we obtain the threshold
\(n \ge K \cdot \ln K\), where
\(K = 8\cdot\delta_1^{-1}\cdot(\Width(\sigma) + \ln|\TTT^*|)\).

\begin{clm}\label{claim:decompose}
  It holds that $\delta_1 \ge \big|\TTT_{\Width(\sigma)}^{\sigma}\big|^{-1} \cdot \big|\TTT_{\Width(\sigma)-1}^{\sigma}\big|$.
\end{clm}
\begin{proof}
We make use of the natural correspondence
\(
  \TTT^{\sigma}_{k} \longrightarrow 
  \prod_{S \subseteq [k]} \partial\TTT^{\sigma}_{|S|},
\)
which expresses every $k$-type as a collection of boundary types over all subsets of variables $x_1,\dots,x_k$.  
This correspondence yields the following estimate:
\begin{flalign}
  \delta_1 &= 
  \prod_{k=1}^{\Width(\sigma)} 
  \big|\partial\TTT^{*}_{k}\big|^{-\binom{\Width(\sigma)-1}{k-1}}
  \ge \prod_{k=1}^{\Width(\sigma)}
  \big|\partial\TTT^{\sigma}_{k}\big|^{-\binom{\Width(\sigma)-1}{k-1}} \nonumber \\
  &= \prod_{k=0}^{\Width(\sigma)}
  \big|\partial\TTT^{\sigma}_{k}\big|^{-\binom{\Width(\sigma)}{k}}
  \times \prod_{k=0}^{\Width(\sigma)-1}
  \big|\partial\TTT^{\sigma}_{k}\big|^{\binom{\Width(\sigma)-1}{k}}
  = \big|\TTT_{\Width(\sigma)}^{\sigma}\big|^{-1} \cdot \big|\TTT_{\Width(\sigma)-1}^{\sigma}\big|.
\end{flalign}
Above, we use that $\binom{\Width(\sigma)-1}{k-1} = \binom{\Width(\sigma)}{k} - \binom{\Width(\sigma)-1}{k}$ for $0 < k < \Width(\sigma)$ to split the product.
\end{proof}

From Claim~\ref{claim:decompose}, we derive the threshold of Lemma~\ref{lemma:algo2-analysis}.
Since
\[
  \Width(\sigma) + \ln|\TTT^*| \le \ln |\TTT^{\sigma}_{\Width(\sigma)}| + \ln\big(\Width(\sigma) \cdot |\TTT^{\sigma}_{\Width(\sigma)}|\big) \le 3 \cdot \ln |\TTT^{\sigma}_{\Width(\sigma)}|,
\]
we have $K \le 24\cdot \delta^{-1}_1\cdot\ln |\TTT^{\sigma}_{\Width(\sigma)}|$.
Therefore, there exists a constant $C > 0$ such that
\begin{equation}\label{eqn:strong-fmp-threshold-lemma}
  K \cdot \ln K \le 24 \cdot \delta_1^{-1} \cdot \underbrace{\ln |\TTT^{\sigma}_{\Width(\sigma)}| \cdot \ln\big(24 \cdot \delta^{-1}_1\cdot\ln |\TTT^{\sigma}_{\Width(\sigma)}| \big)}_{\text{asymptotically dominated by } (\ln |\TTT^{\sigma}_{\Width(\sigma)}|)^2}
  \le C \cdot \frac{|\TTT^{\sigma}_{\Width(\sigma)}|}{|\TTT^{\sigma}_{\Width(\sigma)-1}|} \cdot \big(\ln |\TTT^{\sigma}_{\Width(\sigma)}|\big)^2.
\end{equation}
We conclude that chosing the parameter $n$ to be greater than the right side of~\eqref{eqn:strong-fmp-threshold-lemma} makes the probability of $\fB \models \phi$ at least $1/2$. This establishes Lemma~\ref{lemma:algo2-analysis}. 

\begin{rem}\label{remark:decomposition}
  The threshold of Lemma~\ref{lemma:algo2-analysis} cannot in general be stated in terms of $\TTT^*$, i.e., the set of types appearing in the witness of satisfiability, rather than the full set $\TTT_k^{\sigma}$ of $k$-types over the signature~$\sigma$.  
  The reason is that the correspondence 
  \( \TTT^{\sigma}_{k} \longrightarrow \prod_{S \subseteq [k]} \partial\TTT^{\sigma}_{|S|} \) of Claim~\ref{claim:decompose}
  need not hold when restricted to subsets $\TTT^*_k \subseteq \TTT^{\sigma}_k$. 
  
  For example, suppose $\TTT^*_{2}$ contains only a single $2$-type $\tau$ whose boundary projections on $\{x_1\}$ and $\{x_2\}$ disagree with some $1$-types in~$\TTT^*_{1}$.  
  Then the product decomposition fails, as not every choice of boundary types from $\partial \TTT^*$ can be realised by a $2$-type in $\TTT^*_{2}$.
\end{rem}

\subparagraph{Proof of Theorem~\ref{theorem:precise-bound}.}
To prove the theorem, we first show the following lemma.

\begin{lem}\label{lemma:ratio-types}
  There exists a sequence $(\varepsilon_t)_{t \in \N}$ with $\varepsilon_t \in (0,1/e)$ and $\varepsilon_t \rightarrow 0$
  such that the following holds for every $t \in \N$.
  Let $\sigma = \Rels \uplus \Cons$ be a signature.
  If $\Width(\sigma) \ge t \ge 2$, then
  \[ |\TTT^{\sigma}_{\Width(\sigma)-1}| \ge |\TTT^{\sigma}_{\Width(\sigma)}|^{1/e - \varepsilon_t}. \] 
\end{lem}
\begin{proof}
  Fix $t \ge 2$. Define $\varepsilon_t = e^{-1} - (\frac{t-1}{t})^t$.
  It is known that $(\frac{t-1}{t})^t$ is strictly increasing for all $t \ge 2$ and approaches $e^{-1}$ as $t \rightarrow \infty$.
  In particular, we have $0 < \varepsilon_t < e^{-1}$.

  Let $\sigma$ be the signature with $\Width(\sigma) \ge t$.  
  By Claim~\ref{claim:number-of-types}, the lemma reduces to proving the following inequality:
  \begin{equation}\label{eqn:ratio-types-1}
    \prod_{R \in \Rels} 2^{(\Width(\sigma) + |\Cons| - 1)^{\Arity(R)}} \ge
    \prod_{R \in \Rels} 2^{(e^{-1} - \varepsilon_t) \cdot (\Width(\sigma) + |\Cons|)^{\Arity(R)}}. 
  \end{equation}
  We consider each relation symbol $R \in \Rels$ separately and show that
  \begin{equation}\label{eqn:ratio-types-2}
    2^{(\Width(\sigma) + |\Cons| - 1)^{\Arity(R)}} \ge 2^{(e^{-1} - \varepsilon_t) \cdot (\Width(\sigma) + |\Cons|)^{\Arity(R)}}. 
  \end{equation}
  Taking binary logarithm on both sides of~\eqref{eqn:ratio-types-2} and rearranging yields
  \begin{equation}\label{eqn:ratio-types-3}
    \Big(\frac{\Width(\sigma) + |\Cons| - 1}{\Width(\sigma) + |\Cons|}\Big)^{\Arity(R)} \ge e^{-1} - \varepsilon_t. 
  \end{equation}
  We expand the exponent to get
  \begin{equation}\label{eqn:ratio-types-4}
    \Bigg(\Big(\frac{\Width(\sigma) + |\Cons| - 1}{\Width(\sigma) + |\Cons|}\Big)^{\Width(\sigma) + |\Cons|}\Bigg)^{\Arity(R) / (\Width(\sigma) + |\Cons|)} \ge e^{-1} - \varepsilon_t. 
  \end{equation}
  Since $\Width(\sigma)+|\Cons| \ge t$, inequality~\eqref{eqn:ratio-types-4} is implied by
  \begin{equation}\label{eqn:ratio-types-5}
    \big(e^{-1} - \varepsilon_t \big)^{\Arity(R) / (\Width(\sigma) + |\Cons|)} \ge e^{-1} - \varepsilon_t. 
  \end{equation}
  For $a \eqdef e^{-1} - \varepsilon_t \in (0,1)$ and $\gamma \eqdef \Arity(R) / (\Width(\sigma) + |\Cons|) \in [0,1]$,
  the quantity $a^{\gamma}$ with a fixed $a$ is minimised for $\gamma = 1$. 
  Thus, the left side of~\eqref{eqn:ratio-types-5} is indeed lower bounded by $e^{-1} - \varepsilon_t$, where $\varepsilon_t \in (0,e^{-1})$ and $\varepsilon_t \rightarrow 0$ as $t \rightarrow \infty$, thereby establishing the lemma.
\end{proof}

Having Proposition~\ref{prop:strong-fmp-normal-form} and Lemma~\ref{lemma:ratio-types}, we derive Theorem~\ref{theorem:precise-bound}.
Fix $t \ge 2$.
Let $C > 0$ be as in Proposition~\ref{prop:strong-fmp-normal-form}, and
let $\varepsilon_t \in (0,1/e)$ be as in Lemma~\ref{lemma:ratio-types}.
Choose any $\varepsilon'_t \in (\varepsilon_t,1/e)$ so that $\varepsilon'_t - \varepsilon_t \le 1/t$.
Then there exists a fixed constant $C_t > C$ such that
$C \cdot (\ln x)^2 \le C_t \cdot x^{\varepsilon_t' - \varepsilon_t}$ holds for all $x \ge 1$.
Consequently, for every signature $\sigma$ with $\Width(\sigma) \ge t$, we have
\[  C \cdot \frac{|\TTT^{\sigma}_{\Width(\sigma)}|}{|\TTT^{\sigma}_{\Width(\sigma)-1}|} \cdot \big(\ln |\TTT^{\sigma}_{\Width(\sigma)}|\big)^2 \le C \cdot |\TTT^{\sigma}_{\Width(\sigma)}|^{1-1/e+\varepsilon_t} \cdot \big(\ln |\TTT^{\sigma}_{\Width(\sigma)}|\big)^2 \le C_{t} \cdot |\TTT^{\sigma}_{\Width(\sigma)}|^{1-1/e+\varepsilon'_t}. \]

\section{Tightness Analysis: Enforcing Large Models}
\label{sec:tight}

Setting $\Width(\sigma) = \Theta(|\phi|)$, Proposition~\ref{prop:weak-fmp-normal-form} provides an upper bound of 
\( 2^{2^{\cO(|\phi|\cdot\log|\phi|)}} \) 
on the size of minimal finite models for sentences in $\GF$.  
In this section we show that this bound is essentially tight: we explicitly construct a family of \GF{}-sentences whose minimal models match this doubly exponential upper bound, up to constant factors in the second-level exponent.  
Hence we establish the lower bound of Theorem~\ref{theorem:main}:
for specific sentences $\phi$, models of size $2^{2^{\Omega(|\phi| \cdot \log|\phi|)}}$ are necessary.

\begin{prop}\label{prop:tight}
  There exists a family of \GF{}-sentences $(\phi_n)_{n \in \N}$ such that
  each $\phi_n$ contains neither constants nor equality,
  is over a signature $\sigma_n$ with $\Width(\sigma_n) = n + 4$,
  satisfies $|\phi_n| \le C \cdot n$ for some fixed constant $C > 0$,
  and is satisfiable, but only in domains of size at least \(2^{n!}\).
\end{prop}

The general strategy for enforcing large models is to encode within them a combinatorial object of the desired size.  
A well-known technique for enforcing models of size $2^{2^n}$ is to encode $2^n$-bit numbers from the range $[0,2^{2^n}-1]$.  
However, let us highlight that the standard constructions appearing in the literature (e.g., \cite{Gra99,PH23}) yield formulas of size $\Theta(n^2)$.  
Consequently, they enforce models of size only  
\( 2^{2^{\Omega(\sqrt{|\phi|})}}, \)  
which is asymptotically much smaller than the upper bound \( 2^{2^{\cO(|\phi|\cdot\log|\phi|)}} \) from Proposition~\ref{prop:weak-fmp-normal-form}; even when compared at the doubly logarithmic scale, since $\sqrt{n} / (n\cdot\log n) \rightarrow 0$.
That said, it should be noted that these lower bounds were developed primarily to establish complexity lower bounds, rather than to enforce large domain sizes.

The bottleneck arises from the way these encodings represent $2^n$-bit numbers:
fix two distinct elements $a,b$.
A pair $(a,b)$ is used to encode a $2^n$-bit number $v$.
The $k$-th bit of $v$ is set to $1$ precisely when the atom $Bit(c_1,\dots,c_n)$ holds,  
where the tuple $\langle c_1,\dots,c_n \rangle \in \{ a,b \}^n$ corresponds to the binary representation of~$k$.
Otherwise the $k$-th bit of $v$ is set to $0$.

Yet, specifying that some two sequences $\langle c_1,\dots,c_n \rangle$ and $\langle c_1',\dots,c_n' \rangle$ correspond to respectively the $k$-th and $(k{+}1)$-st bits of $v$ requires examining $n$ different cases to account for potential carries, which forces formulas of size $\Omega(n^2)$.  

To reach the optimal lower bound of  
\( 2^{2^{\Omega(|\phi|\cdot \log |\phi|)}} \)  
a different construction is required.  
Our approach is to encode the powerset of the set of all permutations of $n$ elements, which has cardinality  
\( 2^{n!} = 2^{2^{\Theta(n \cdot \log n)}}. \)
We use a certain decomposition of permutations from~\cite{FKM24} to achieve this with formulas of length $\cO(n)$, 
thereby ensuring the desired growth rate.

\subparagraph{Proof of Proposition~\ref{prop:tight}.}
Fix an integer \(n\ge 3\).  We shall construct a \(\GF\)-sentence \(\varphi_n\) designed to encode the powerset of \(S_n\) (the set of permutations of \(\{1,\dots,n\}\)).
The signature of \(\varphi_n\) contains the following relation symbols:
\begin{equation}\label{eqn:phi-signature}
  \sigma_n \;=\; \{\,Wit,\,Mem,\,Adj,\,Gen,\,Inc,\,Dec,\,Succ,\,P_1,\dots,P_n,\,Q_1,\dots,Q_{n-1}\,\},
\end{equation}
where the arity of $Wit$ and $Mem$ is $n+1$; of $Adj$ is $n+2$; of $Gen$, $Inc$, and $Dec$ is $n+3$; of $Succ$ is $n+4$; and the symbols $P_i$ and $Q_i$ are all unary.
We write \(\xs=\langle x_1,\dots,x_n \rangle\) and \(\ys=\langle y_1,y_2 \rangle\) for the short-hand tuples that occur repeatedly below.  Let \(\pi_{\mathrm{cyc}}\) denote the cyclic permutation \(1\mapsto 2\mapsto\cdots\mapsto n\mapsto 1\), and let \(\pi_{\mathrm{swp}}\) denote the transposition swapping \(1\) and \(2\).

The construction relies on encoding subsets of the symmetric group \(S_n\).  
We begin with a fixed ``ground set'' of size \(n\), represented by some \(n\)-tuple of elements, denoted \(\as\).  

For every subset \(F \subseteq S_n\) we postulate the existence of a \emph{witness element} \(w_F\).  
The witness \(w_F\) is linked uniformly to all permutations of the ground set: for each \(\pi \in S_n\) we require  
\(Wit(\pi(\as),w_F)\).
The actual membership of a permutation \(\pi\) in \(F\) is then expressed via the predicate \(Mem\):  
\(Mem(\pi(\as),w_F)\) if and only if \(\pi \in F\).
This forces distinct subsets \(F,F'\) to be represented by distinct witnesses \(w_F \neq w_{F'}\).  
Thus, if we succeed in enforcing the existence of witnesses for all subsets of \(S_n\), the model will contain at least \(2^{n!}\) distinct elements.  

To navigate between different subsets, observe that one can transform any subset \(F \subseteq S_n\) into any other \(F' \subseteq S_n\) by toggling permutations one by one.  
We capture this stepwise modification using the relation \(Adj\): the atom  
\(Adj(\pi(\as),w_F,w_{F'})\)
is intended to hold precisely when the symmetric difference of \(F\) and \(F'\) is the singleton \(\{\pi\}\).  

To axiomatise the relation $Adj$, we employ a certain decomposition of permutations,
which will be implemented using the remaining auxiliary predicates.
Details of this decomposition and its encoding will be presented below.

We first postulate the existence of the ground set and an initial witness:
\begin{equation}\label{phi-first}
  \exists \xs,y~\Big(Wit(\xs,y) \wedge \bigwedge_{i=1}^{n} P_i(x_i)\Big).
\end{equation}
In what follows, we denote the tuple witnessing for $\xs$ by \(\as = \langle a_1,\dots,a_n \rangle \).
We require the components of \(\as\) to be pairwise distinct.
For this we axiomatise that $\neg P_i(x) \vee \neg P_j(x)$ holds for distinct $i \neq j$.
However, the obvious sentence would be of size $\Theta(n^2)$. To keep the formula of size linear in $n$,
we use the auxiliary predicates $Q_1,\dots,Q_{n-1}$:
\begin{equation}
  \bigwedge_{i=2}^{n} \forall x\;\big(P_i(x) \rightarrow Q_{i-1}(x)\big)
  \wedge \bigwedge_{i=2}^{n-1} \forall x\;\big(Q_i(x) \rightarrow Q_{i-1}(x)\big)
  \wedge \bigwedge_{i=1}^{n-1} \forall x\;\big(Q_i(x) \rightarrow \neg P_i(x)\big).
\end{equation}

Next, we ensure that witnesses respect the intended invariant:  
for every witness $y$ and
every permutation \(\pi \in S_n\) we must have \(Wit(\pi(\as),y)\).  
Since \(S_n\) is generated by \(\pi_{\rm cyc}\) and \(\pi_{\rm swp}\), it suffices to enforce closure under these two transformations:
\begin{equation}\label{phi-perms}
  \forall \xs,y~\big(Wit(\xs,y) \rightarrow
  \big(Wit(\pi_{\rm swp}(\xs),y) \wedge Wit(\pi_{\rm cyc}(\xs),y)\big)\big).
\end{equation}

We now enforce that every set \(F\) can be extended to a set \(F'\) differing precisely on a single permutation \(\pi\).  
Formally, we require that for each permutation tuple \(\pi(\as)\) and each witness \(y_1\), there exists another witness \(y_2\) that differs exactly on the membership of \(\pi\):
\begin{equation}\label{phi-adj}
  \forall \xs,y_1~
  \big(Wit(\xs,y_1) \rightarrow \exists y_2~
  \big(Adj(\xs,y_1,y_2) \wedge Wit(\xs,y_2) \wedge
  \big(Mem(\xs,y_1) \oplus Mem(\xs,y_2)\big)\big)\big).
\end{equation}

To axiomatise that the witnesses $y_1$ and $y_2$ agree on every other permutation $\pi' \neq \pi$, i.e., $Mem(\pi'(a),y_1) \leftrightarrow Mem(\pi'(a),y_2)$, we employ the following combinatorial lemma.
\begin{lem}[Lemma 7.2 in \cite{FKM24}]\label{l:permdistinct}
  Let $\pi,\pi' \in S_n$ be permutations. 
  Then $\pi' \neq \pi$ if and only if 
  \(
    \pi' = \pi_{\rm cyc}^{-j}\circ \rho \circ \pi_{\rm cyc}^k\circ\pi
  \)
  for some $0\leq j < k < n$ and some $\rho \in S_n$ fixing $n$ (i.e., $\rho(n)=n$).
\end{lem}
\begin{proof}
  By the group action $x \mapsto x \circ \pi^{-1}$, assume that $\pi = \ONE$ is the identity permutation.

  First, suppose $\pi'\neq \ONE$.
  Necessarily, there exists $i\in[n]$ such that $\pi'(i) > i$.
  We set $k \eqdef n-i$ and~$j \eqdef n-\pi'(i)$, so that $0\leq j<k<n$.
  Define~$\rho \eqdef \pi_{\rm cyc}^{j} \circ \pi' \circ \pi_{\rm cyc}^{-k}$.
  By definition, it holds that~$\pi' = \pi_{\rm cyc}^{-j}\circ \rho\circ\pi_{\rm cyc}^k$.
  We verify that~$\rho$ fixes the position $n$:
  \begin{flalign*}
    \rho(n)
    =(\pi_{\rm cyc}^{j} \circ \pi' \circ \pi_{\rm cyc}^{-k})(n)
    =(\pi_{\rm cyc}^j \circ \pi' \circ \pi_{\rm cyc}^{i})(n)
    =(\pi_{\rm cyc}^j\circ\pi')(i)
    =\pi_{\rm cyc}^{-\pi'(i)}\big(\pi'(i)\big)
    =n.
  \end{flalign*}

  Now, we show that the conditions of the lemma holding for $\pi' = \ONE$ imply $\pi_{\rm cyc}^{-j}\circ \rho \circ \pi_{\rm cyc}^k \neq \ONE$.
  Since $j,k \in [0,n{-}1]$ and $\rho(n) = n$, we can have $\pi_{\rm cyc}^{j - k} = \rho$ only for $j = k$.
  However, the lemma requires $j < k$, and therefore $\pi_{\rm cyc}^{-j}\circ \rho \circ \pi_{\rm cyc}^k \neq \ONE$.
\end{proof}

We implement the decomposition of Lemma~\ref{l:permdistinct} using the auxiliary predicates 
\(Gen\), \(Inc\), \(Dec\), and \(Succ\).  
Intuitively:  
\(Inc\) increases the counter along powers of $\pi_{\rm cyc}$;  
\(Gen\) introduces the action of $\rho$ restricted to $[n{-}1]$, generated by $\pi_{\rm swp}$ and $\pi'_{\rm cyc}$: $1\mapsto 2\mapsto\cdots\mapsto n{-}1\mapsto 1$;  
\(Dec\) decreases the counter via powers of $\pi_{\rm cyc}^{-1}$;  
and \(Succ\) provides the successor relation on the counter elements.  
The counter itself is represented by the last argument, where an element marked by $P_i$ with $1 \le i \le n-1$ indicates the difference $i \eqdef k - j$ ($j$, $k$ as in Lemma~\ref{l:permdistinct}).
This ensures that we never reach $0$ or $n$, thereby avoiding the trivial case $\pi'=\pi$.

We axiomatise \(Succ\) as:
\begin{gather}\label{phi-counter}
  \forall \xs,\ys,z,z'~
  \big(Succ(\xs,\ys,z,z') \rightarrow \bigvee_{i=1}^{n-2} \big(P_i(z) \wedge P_{i+1}(z')\big)\big).
\end{gather}
Next, we generate permutations of the form $\pi_{\rm cyc}^k\circ\pi$ for $0<k<n$:
\begin{gather}
  \forall \xs,\ys~
  \big(Adj(\xs,\ys) \rightarrow \exists z~\big(Inc(\pi_{\rm cyc}(\xs),\ys,z) \wedge P_1(z)\big)\big), \\
  \forall \xs,\ys,z~
  \big(Inc(\xs,\ys,z) \rightarrow \big(P_{n-1}(z) \vee
  \exists z'~\big(Succ(\xs,\ys,z,z') \wedge Inc(\pi_{\rm cyc}(\xs),\ys,z')\big)\big)\big).
\end{gather}
Then, we generate permutations of the form $\rho\circ\pi_{\rm cyc}^k\circ\pi$, where $\rho$ fixes $n$.  
The subgroup $\{ \rho \in S_n \mid \rho(n) = n \}$ is induced by $\pi_{\rm swp}$ and $\pi'_{\rm cyc}$: $1\mapsto 2\mapsto\cdots\mapsto n{-}1\mapsto 1$.
We enforce:
\begin{gather}
  \forall \xs,\ys,z~\big(Inc(\xs,\ys,z) \rightarrow Gen(\xs,\ys,z)\big), \\
  \forall \xs,\ys,z~
  \big(Gen(\xs,\ys,z) \rightarrow
  \big(Gen(\pi'_{\rm cyc}(\xs),\ys,z) \wedge Gen(\pi_{\rm swp}(\xs),\ys,z)\big)\big).
\end{gather}
Finally, we generate permutations of the form $\pi_{\rm cyc}^{-j}\circ\rho\circ\pi_{\rm cyc}^k\circ\pi$ with $0 \le j<k < n$:
\begin{gather}
  \forall \xs,\ys,z~\big(Gen(\xs,\ys,z) \rightarrow Dec(\xs,\ys,z)\big), \\
  \forall \xs,\ys,z~
  \big(Dec(\xs,\ys,z) \rightarrow \big(P_1(z) \vee
  \exists z'~\big(Succ(\xs,\ys,z',z) \wedge Dec(\pi_{\rm cyc}^{-1}(\xs),\ys,z')\big)\big)\big).
\end{gather}
By Lemma~\ref{l:permdistinct}, every $\pi'\neq\pi$ admits a representation such that 
$Dec(\pi'(\as),y_1,y_2,z)$ holds.
We then enforce membership agreement between $y_1$ and $y_2$ on all such permutations:
\begin{gather}
  \forall \xs,y_1,y_2,z~
  \big(Dec(\xs,y_1,y_2,z) \rightarrow \big(Mem(\xs,y_1) \leftrightarrow Mem(\xs,y_2)\big)\big). \label{phi-last}
\end{gather}

We define $\varphi_n$ to be the conjunction of the sentences~(\ref{phi-first})--(\ref{phi-last}).
Since each conjunct has length $\cO(n)$, the resulting sentence $\varphi_n$ satisfies
$|\varphi_n| = \cO(n)$.
Moreover, the signature $\sigma_n$ of $\varphi_n$ has width $n+4$, as required.
Hence the syntactic assumptions of Proposition~\ref{prop:tight} are met.

In what follows, we prove Claims~\ref{claim:phi-models-standard}
and~\ref{claim:phi-models-lowerbound}, which establish, respectively,
that $\varphi_n$ is satisfiable and that every model of $\varphi_n$
has domain size at least $2^{n!}$.  
This completes the proof of Proposition~\ref{prop:tight}.

\begin{clm}\label{claim:phi-models-standard}
  The sentence \(\varphi_n\) is satisfiable: there is a standard model \(\fA_n\models\varphi_n\).
\end{clm}
\begin{proof}
  Define the standard model \(\fA_n\) as follows. Take the ground tuple \(\as=\langle 1,\dots,n \rangle\),
  and for every $i \in [n]$ mark the position $i$ with the unary predicates $P_i$ and $Q_1,\dots,Q_{i-1}$.
  Then for every subset \(F\subseteq S_n\) introduce a distinct element \(w_F\).
  Interpret
  \[
    Wit^{\fA_n}=\{\langle \pi(\as),w_F \rangle \mid \pi\in S_n,\ F\subseteq S_n\},
    \quad
    Mem^{\fA_n}=\{\langle \pi(\as),w_F \rangle \mid F \subseteq S_n,\ \pi\in F\},
  \]
  and
  \[
    Adj^{\fA_n}=\{\langle \pi(\as),w_F,w_{F'} \rangle \mid \pi \in S_n,\ F,F' \subseteq S_n,\ F\triangle F'=\{\pi\}\}.
  \]
  The remaining predicates $Inc$, $Dec$, $Gen$, and $Succ$ are interpreted so as to implement
  the decomposition from Lemma~\ref{l:permdistinct} (this is straightforward and was sketched above).
  By construction all conjuncts (\ref{phi-first})--(\ref{phi-last}) are satisfied, hence \(\fA_n\models\varphi_n\).
\end{proof}

\begin{clm}\label{claim:phi-models-lowerbound}
  In every model \(\fB\models\varphi_n\) there are at least \(2^{n!}\) distinct elements.
\end{clm}
\begin{proof}
  Let \(\fB\models\varphi_n\).
  Fix a tuple \(\as\) witnessing $\xs$ in~\eqref{phi-first} to serve as the ground set.
  Define a mapping that assigns to each subset \(F\subseteq S_n\) an element \(w_F\) such that the following holds:
  \begin{enumerate}
    \item for every \(\pi\in S_n\), \(\fB \models Wit(\pi(\as),w_F)\), and
    \item for every \(\pi\in S_n\), \(\fB \models Mem(\pi(\as),w_F)\) iff \(\pi\in F\).
  \end{enumerate}
  Such a mapping $F \mapsto w_F$ exists and is necessarily injective.
  
  Indeed:
  The axiom~\eqref{phi-first} postulates the existence of the ground set $\as$ together with an initial witness $w_{F_0}$ for some subset $F_0 \subseteq S_n$.
  The axioms~\eqref{phi-perms}--\eqref{phi-adj} express the existence of all $Adj$-neighbours:
  for every witness $w_F$ and every permutation $\pi \in S_n$, there exists an $Adj$-neighbour of $w_F$ differing on \(\pi\).
  The axioms~\eqref{phi-counter}--\eqref{phi-last} together with Lemma~\ref{l:permdistinct} force the intended interpretation of the $Adj$ predicate: traversing along $Adj$-connection toggles precisely this single permutation \(\pi\)
  while preserving all other memberships.
  Hence, we can find $w_F$ for every $F \subseteq S_n$, and moreover $w_F = w_{F'}$ for two distinct subsets $F$ and $F'$ would produce a contradiction.

  Therefore the witnesses \(w_F\) all exist and are pairwise distinct for distinct \(F\subseteq S_n\).
  Since there are \(2^{|S_n|}=2^{n!}\) subsets of \(S_n\), the domain of \(\fB\) contains at least \(2^{n!}\) distinct elements.
\end{proof}

\subparagraph{Optimality of Theorem~\ref{theorem:precise-bound}.}
Theorem~\ref{theorem:precise-bound} bounds the minimal model size of a normal-form \GF{}-sentence~$\phi$ in terms of the number of $k$-types over its signature~$\sigma$: informally, if $\phi$ is satisfiable, then it has a model of size $\cO\big(|\TTT^{\sigma}_{\Width(\sigma)}|^{0.63}\big)$.
This formulation, however, does not constrain the length of~$\phi$.
We now show that a comparable model size can already be enforced---albeit only at the doubly logarithmic scale---by sentences of linear length in~$\Width(\sigma)$.

Corollary~\ref{corollary:precise-tight} refines the perspective of Proposition~\ref{prop:tight}:
for every $k \in \N$,
there exists a sentence of length~$\cO(k)$ enforcing models of size comparable to the number of all $k$-types over the underlying signature of width $k$, up to an arbitrarily small loss at the doubly logarithmic scale.
Beyond guaranteeing large domain size, it also shows that such sentences enforce the realisation of a substantially large set of $k$-types, in the same asymptotic regime.

\begin{cor}\label{corollary:precise-tight}
  There exists a family of \GF{}-sentences 
  $(\phi_n)_{n \in \N}$ as in Proposition~\ref{prop:tight} such that the following holds.
  For every $\varepsilon > 0$ there exists a threshold $n_{\varepsilon} \in \N$ such that,
  whenever $n \ge n_{\varepsilon}$ and $\fB$ is a model of $\phi_n$, we have
  \[
    \log_2\log_2|B| \ge (1-\varepsilon) \cdot \log_2\log_2 |\TTT^{\sigma_n}_{\Width(\sigma_n)}|
    \;\;\text{and}\;\;
    \log_2\log_2|\TTT^{\fB}_{\Width(\sigma_n)}| \ge (1-\varepsilon)\cdot\log_2\log_2 |\TTT^{\sigma_n}_{\Width(\sigma_n)}|.
  \]
\end{cor}
\begin{proof}
  Let $(\phi_n)_{n \in \N}$ be a family of sentences as constructed in the proof of Proposition~\ref{prop:tight}.

  Let $n \ge 3$, and let $\fB$ be a model of $\varphi_n$. 
  We observe that $\fB$ has not only a domain of size at least $2^{n!}$ but also realises at least $2^{n!}$ distinct $(n+4)$-types:
  following the proof of Claim~\ref{claim:phi-models-lowerbound},
  select a tuple $\as$ and for each $F \subseteq S_n$ choose a witness $w_F$.
  Consider the set
  \[ \big\{ \, \type{\fB}{\as,w_{F_1},w_{F_2},w_{F_3},w_{F_4}} \bigmid w_{F_1},\dots,w_{F_4} \in B \setminus \as \;\text{are pairwise distinct} \, \big\}. \]
  The size of this set is at least $(2^{n!} - n - 3)^4 \ge 2^{n!}$,
  since each type $\type{\fB}{\as,w_{F_1},w_{F_2},w_{F_3},w_{F_4}}$ determines the subsets $F_1,\dots,F_4 \subseteq S_n$ via the predicate $Mem$.
  Therefore $|\TTT^{\fB}_{\Width(\sigma_n)}| \ge 2^{n!}$.

  Fix $\varepsilon > 0$.
  To prove the claim, it remains to show that
  $\log_2 n! \ge (1-\varepsilon) \cdot \log_2\log_2|\TTT^{\sigma_n}_{\Width(\sigma_n)}|$
  holds for all $n \ge n_{\varepsilon}$, where $n_{\varepsilon} \in \N$ is some threshold depending solely on the chosen $\varepsilon$.

  Since $\Width(\sigma_n) = n + 4$ and $|\sigma_n| = 2 \cdot n + 6$,
  from Claim~\ref{claim:number-of-types} we have
  \begin{equation}
   \log_2 |\TTT^{\sigma_n}_{\Width(\sigma_n)}|
    \le |\sigma_n| \cdot \Width(\sigma_n)^{\Width(\sigma_n)}
    \le (2 \cdot n + 6) \cdot (n+4)^{n+4}
    \le (n+6)^{n+6}.
  \end{equation}
  Using Stirling's bound $n! \ge (n/e)^n$
  and the fact that $n/e \ge (n+6)^{1-\varepsilon/2}$ for all sufficiently large $n \in \N$,
  we obtain that there exists $n_{\varepsilon} \in \N$ such that, whenever $n \ge n_{\varepsilon}$,
  \begin{equation}
    n! \ge (n/e)^n
    \ge (n+6)^{(1-\varepsilon/2)\cdot n}
    \ge (n+6)^{(1-\varepsilon) \cdot (n+6)}
    \ge \big(\log_2 |\TTT^{\sigma_n}_{\Width(\sigma_n)}|\big)^{1-\varepsilon}.
  \end{equation}
\end{proof}

\section{Extension to a Stronger Fragment: the Triguarded Fragment}
\label{sec:beyond}

The \emph{Triguarded Fragment} (\TGF{}) extends the Guarded Fragment by admitting one additional rule in the syntax of \GF{}  (Definition~\ref{definition:GF-syntax}):
\begin{enumerate}[label=(\roman*)]\addtocounter{enumi}{3}
  \item\label{TGF-rule}
    Let $x,y$ be variables and let $\psi$ be a formula in \TGF{}.
    If $\FreeVars(\psi) \subseteq \{x,y\}$, then both $\exists x\,\psi$ and $\forall x\,\psi$ belong to \TGF{}.
\end{enumerate}

Rule~\ref{TGF-rule} relaxes the guarding requirement of \GF{} in the case of formulas with at most two free variables.
Thus, \TGF{} permits unguarded quantification over pairs of elements.
To compare, recall that \GF{} allows for free quantification only over individual elements.

In this section we establish Theorem~\ref{theorem:TGF-main}, showing that the equality-free subfragment of \TGF{} enjoys the finite model property with a doubly exponential upper bound on model size.  

\subparagraph{Proof of Theorem~\ref{theorem:TGF-main}.}
To unify the syntax of \GF{} and \TGF{}, it is convenient (following~\cite{RS18}) to introduce an auxiliary fragment, denoted \GFU{}.  
This is simply \GF{} over signatures extended by a \emph{distinguished} binary symbol $U$, interpreted as the full binary relation on the domain.  
In the terminology of description logics, $U$ is referred to as a \emph{universal role}.
With this convention, rule~\ref{TGF-rule} can be viewed as a special case of the guarded quantifier rule~\ref{GF-rule3}, 
with $U$ acting as a dummy guard for formulas with at most two free variables.  
In particular, Theorem~\ref{theorem:TGF-main} is obtained as a corollary of
Theorem~\ref{theorem:main}, in combination with the following lemma.

\begin{lem}\label{lemma:GFU}
  Let $\TTT^* = \bigcup_{k=0}^{\Width(\sigma)} \TTT^*_k$ be a satisfiability witness, and let $n \in \N$ be arbitrary.
  Assume that $\TTT^*$ satisfies the following condition:
  \begin{align*}
    &\text{for every pair of $1$-types $\tau_1, \tau_2 \in \TTT^*_1$, 
  there exists a $2$-type $\tau_{1,2} \in \TTT^*_2$ such that} \\
  &\text{its endpoints are $\tau_1$ and $\tau_2$, that is, $\type{\tau_{1,2}}{x_1} = \tau_1$ and $\type{\tau_{1,2}}{x_2} = \tau_2$.}
  \end{align*}
  Then Algorithm~\ref{algos2} given $\TTT^*$ and $n$ generates the structure $\fB$ with the property that between every pair of elements the $2$-type is defined, i.e., \(\type{\fB}{a,b} \in \TTT^*_2\) for all distinct $a,b \in [n]$.
\end{lem}

The constraint on the satisfiability witness $\TTT^*$ in Lemma~\ref{lemma:GFU} guarantees that the symbol $U$ is interpreted as a universal relation: 
it ensures that Algorithm~\ref{algos2} always has at least one compatible $2$-type to assign to every pair of elements.\footnote{Note that this property does not hold for Algorithm~\ref{algos},  
since it selects $2$-types from the entire space rather than restricting to those compatible with the already assigned $1$-types.}

The model construction for \GFU{} is clearly \emph{sound}: given a witness $\TTT^*$ satisfying the constraint of Lemma~\ref{lemma:GFU}, the algorithm produces a valid model $\fB$. 
A natural question is whether this condition is also \emph{complete}—i.e., whether for every satisfiable \GFU{}-sentence
we can always find a model $\fA$ of $\phi$ and extract a witness $\TTT^*$ from $\fA$ that satisfies this constraint.

The sufficient condition for completeness is straightforward: 
the only problematic case arises when both $1$-types are identical.  
Hence the requirement reduces to ensuring that, for every $1$-type realised in a model, 
there exist at least two distinct unnamed elements realising it.\footnote{Recall that, by our (non-standard) convention, $1$-types are defined only for unnamed elements.}    
A standard argument shows that such models always exist for equality-free sentences:
\begin{lem}[6.2.26 in~\cite{BGG}]\label{claim:duplicate}
  Let $\sigma = \Rels \uplus \Cons$ be a signature.
  Let $\fA$ be a $\sigma$-structure with domain $A = A_0 \uplus \Cons$.
  Define $2 \ltimes \fA$ as the $\sigma$-structure with domain 
  $\{0,1\} \times A$, 
  where each constant $c \in \Cons$ is identified with $(c,0)$, 
  and for every $k$-ary symbol $R \in \Rels$ we set
  \[
    R^{2\ltimes\fA} = \bigcup\nolimits_{i_1,\dots,i_k \in \{0,1\}} 
    \big\{\,\langle (i_1,a_1),\dots,(i_k,a_k) \rangle \bigmid \langle a_1,\dots,a_k \rangle \in R^\fA \,\big\}.
  \]
  Then $\fA$ and $2 \ltimes \fA$ satisfy exactly the same equality-free sentences.
\end{lem}

Consequently, by the reduction to the equality-free \GFU{}, this establishes Theorem~\ref{theorem:TGF-main}.

\subparagraph{Remarks on Equality.}
We conclude this section by considering a semantic constraint that allows equality in Theorem~\ref{theorem:TGF-main}.
A similar observation has already appeared in Gurevich and Shelah's work concerning the probabilistic proof for the G\"odel class~\cite{GS83}.

\smallskip
The Triguarded Fragment with equality is undecidable~\cite{RS18}. 
However, nowhere in our argument did we explicitly exclude the use of equality, except the final argument in Lemma~\ref{claim:duplicate}.  
This observation allows us to recover decidability of \TGF{} with equality under the following semantic constraint:  
every $1$-type realised by an unnamed element in a model must have at least two distinct unnamed realisations.  

Readers familiar with Gr\"adel, Kolaitis, and Vardi's proof of the finite model property for \FOt{}~\cite{GKV97} may recall a related notion, where elements realising unique $1$-types are called \emph{kings}.  
In our setting, a \emph{king} is an unnamed element whose $1$-type is distinct from $1$-types of all other unnamed elements. 
Naturally, under the above semantic constraint, kings are excluded.  
This allows us to strengthen Theorem~\ref{theorem:TGF-main}, and as we discuss below, also Theorem~\ref{theorem:main}.  

\begin{cor}\label{corollary:TGF}
  Let $\phi$ be a \TGF{}-sentence.  
  If $\phi$ has a model without kings, then it also has a finite model, 
  whose domain size satisfies the same bound as in Theorem~\ref{theorem:TGF-main}.
\end{cor}

At first sight, banning kings may appear to be an artificial restriction.
However, we believe that it has natural applications.  
For example, in the theory of directed graphs with a single binary predicate $E$, tournament graphs are modelled by
$\forall x\, \neg E(x,x)$ and $\forall x,y\, \big(x \neq y \rightarrow \big(E(x,y) \leftrightarrow \neg E(y,x)\big)\big)$.
This property of graphs is expressible neither in \GF{} (with equality) nor in the equality-free \TGF{}.
Naturally, non-singleton tournament graphs satisfy the restriction on kings: every vertex realises the same $1$-type $\{ \neg E(x_1,x_1) \}$.

Moreover, the constraint of Corollary~\ref{corollary:TGF} is vacuous for \GF{} (even with equality): take two disjoint copies of a model and glue them on constants. Standard arguments show that the resulting structure satisfies the same guarded sentences.
Consequently, \TGF{} without kings strictly subsumes---at least from the perspective of satisfiability---not only the equality-free \TGF{} but also \GF{} (with equality).

Further applications come from description logics (DLs), which are often studied as fragments of \GF{} via the so-called \emph{standard translation}.
Corollary~\ref{corollary:TGF} does not limit the use of constants: atoms of the form $x=c$ or $c = c'$ for constants $c$ and $c'$ pose no problem.
In the standard translation, such atoms are often used to capture DLs with \emph{nominals}.

Last but not least, constants can often be used to \emph{simulate} kings.  
By introducing a fresh constant for each king in a model, one can transform the model into a king-free one.
Such a transformation can even serve as a basis for a practical procedure semi-deciding the finite satisfiability problem for the full \TGF{}---namely, a sentence of \TGF{} is finitely satisfiable iff it has a king-free model over a signature expanded by $k$ fresh constants for some $k \in \N$.
As Example~\ref{example:remove-kings} witnesses, this does not contradict the general undecidability of \TGF{}.  

\begin{exa}\label{example:remove-kings}
  Consider the structure $\fN$ over the natural numbers $\N$, interpreting a binary relation 
  \(
  Succ^{\fN} = \{ \langle n, n+1 \rangle \mid n \in \N \}
  \)
  and a unary predicate 
  \(
  Zero^{\fN} = \{0\}.
  \)
  In $\fN$, the element $0$ is clearly a king: it is the only element whose $1$-type contains the atom $Zero(x)$.

  Now take any finite set $K \subseteq \N$, and expand $\fN$ to a structure $\fM$ by interpreting, for each $k \in K$, a new constant $k$ as the element $k$ itself.  
  However, $\fM$ again contains a king: let $m$ be the maximal element of $K$.  
  Then the element $m+1$ has a unique $1$-type, since it is the only element for which the atom $Succ(m,x)$ holds.  
  Thus, adding finitely many constants does not in general eliminate kings.  

  The duplication trick from Lemma~\ref{claim:duplicate} does not resolve the issue either:
  the \TGF{}-sentence \(\forall x,y \;\big(\big(Zero(x) \wedge Zero(y)\big) \rightarrow x = y\big)\) is no longer satisfied in the structure $2 \ltimes \fN$.

  For a formal proof that kings cannot be eliminated in general, see Goldfarb’s undecidability result for the G\"odel class with equality~\cite{Gol84}, which embeds into \TGF{}.
\end{exa}

\section{Derandomisation: an Explicit Construction of Models}
\label{sec:algebra}

In Section~\ref{sec:random} using probabilistic arguments we established Theorem~\ref{theorem:main}: every satisfiable sentence $\phi$ of \GF{} admits a finite model of doubly exponential size.
However, the probabilistic proof is inherently non-constructive: it does not produce a concrete model of $\phi$, but instead shows only that models of $\phi$ constitute a non-zero fraction of all structures of the given size.

We now turn to a constructive variant of Theorem~\ref{theorem:main},  
previously stated without details in Proposition~\ref{prop:constructive}.  
With the necessary tools in place, we give a formal version:

\begin{prop}\label{prop:det-fmp-normal-form}
  There exists a deterministic algorithm and a constant $C > 0$ such that the following holds.  
  For every normal-form \GF-sentence $\phi$ over a signature $\sigma$,  
  given a witness of satisfiability $\TTT^* = \bigcup_{k=0}^{\Width(\sigma)} \TTT^*_k$ for $\phi$  
  and a parameter $n \in \N$ satisfying
  \[
    n \ge  C \cdot |\TTT^{\sigma}_{\Width(\sigma)}|^{2^{\Width(\sigma)}},
  \]
  the algorithm constructs a structure $\fB$ with unnamed domain of size $n$ such that $\fB \models \phi$.
  The algorithm runs in time polynomial in $n^{\Width(\sigma)}$ and the encoding size of $\TTT^*$.
\end{prop}

Our strategy is to derandomise the probabilistic proof  
by giving explicit values for the random choices in Algorithm~\ref{algos}.  
The resulting deterministic procedure (Algorithm~\ref{det-algos})  
relies on algebraic hash functions to simulate randomness,  
and we prove that with appropriate parameters, it always produces a valid model (Lemma~\ref{lemma:det-algo-analysis}).

\begin{algorithm}[t]
  \caption{Deterministic model generation from a satisfiability witness}\label{det-algos}
  \KwIn{signature $\sigma = \Rels \uplus \Cons$;  
        satisfiability witness $\{\TTT^*_k\}_{k=0}^{\Width(\sigma)}$;  
        hash~parameters~$p_1,\dots,p_{2^{\Width(\sigma)}-1} \in \N$; number $M \eqdef \prod_{i=1}^{2^{\Width(\sigma)}-1} p_i$}
  \KwOut{$\sigma$-structure $\fB$ with domain $B \eqdef [0,\Width(\sigma)-1] \times [0,M-1] \uplus \Cons$}
  initialise $\fB$ as the empty $\sigma$-structure with domain $B$\;
  assign the unique $0$-type from $\TTT^*_0$ to $\fB$\;
  \For{$k = 1,2,\dots,\Width(\sigma)$\label{det-line:outer}}{
    let $\TTT^*_k(0),\dots,\TTT^*_k(|\TTT^{*}_k|-1)$ enumerate $\TTT^{*}_k$\label{det-line:enum}\;
    \ForEach{$(\alpha_1,\beta_1),\dots,(\alpha_k,\beta_k) \in [0,\Width(\sigma){-}1] \times [0,M{-}1]$ \rm{with} $\alpha_1 < \cdots < \alpha_k$\label{det-line:inner}}{
      compute index $i \eqdef 2^{\alpha_1} + \dots + 2^{\alpha_k}$\label{det-line:index}\;
      compute hash $h \eqdef ((\beta_1 + \dots + \beta_k) \bmod p_i) \bmod |\TTT^{*}_k|$\label{det-line:hash}\;
      let $\tau \eqdef \TTT^*_k(h)$\label{det-line:choose}\;
      \If{$\tau$ is compatible with $\type{\fB}{(\alpha_1,\beta_1),\dots,(\alpha_k,\beta_k)}$\label{det-line:test}}{
        assign $\type{\fB}{(\alpha_1,\beta_1),\dots,(\alpha_k,\beta_k)} \eqdef \tau$\;
      }
    }
  }
\end{algorithm}

\begin{lem}\label{lemma:det-algo-analysis}
  Let $\sigma$ be the signature of $\phi$, and let $\TTT^* = \bigcup_{k=0}^{\Width(\sigma)} \TTT^*_k$ be a satisfiability witness for $\phi$.  
  Suppose the hash parameters $p_1,\dots,p_{2^{\Width(\sigma)}-1}$ are chosen so that
  \begin{enumerate}
    \item $p_1,\dots,p_{2^{\Width(\sigma)}-1}$ are all prime, and
    \item \(|\TTT^{\sigma}_{\Width(\sigma)}| \le p_1 < \dots < p_{2^{\Width(\sigma)}-1}\).
  \end{enumerate}
  Then the structure $\fB$ generated by Algorithm~\ref{det-algos} satisfies $\fB \models \phi$.
\end{lem}

\subparagraph{Proof of Lemma~\ref{lemma:det-algo-analysis}.}
Let us assume that every type in $\TTT^*$ is guarded.
This assumption is used in Claim~\ref{claim:det-tau-extension}, and can be later removed, see Remark~\ref{remark:dense-types}. 

Set $M = \prod_{i=1}^{2^{\Width(\sigma)}-1} p_i$.
For each $k \in [\Width(\sigma)]$, let $\Omega_k$ denote the set of $k$-tuples processed in the inner loop at Line~\ref{det-line:inner}, during the $k$th iteration of the outer loop at Line~\ref{det-line:outer}:
\[
  \Omega_k = \big\{ \langle (\alpha_1,\beta_1),\dots,(\alpha_k,\beta_k) \rangle \bigmid 
      0 \le \alpha_1 < \dots < \alpha_k < \Width(\sigma),\;
      \beta_1,\dots,\beta_k \in [0,M-1] \big\}.
\]
For technical reasons, we also declare $\Omega_0$ as the set with the unique zero-length tuple.

For each $k \in [\Width(\sigma)]$ and each $k$-tuple $\bs \in \Omega_k$,  
let $\cX(\bs)$ denote the $k$-type $\tau$ chosen by Algorithm~\ref{det-algos} at Line~\ref{det-line:choose} when processing $\bs$.  
For a permutation $\rho\colon [k] \rightarrow [k]$, we naturally write
\(\cX(\rho(\bs)) = \type{\cX(\bs)}{x_{\rho(1)},\dots,x_{\rho(k)}}, \)
that is, the $k$-type $\cX(\bs)$ reindexed according to $\rho$.

The values $\cX(\bs)$ can be viewed as a particular instantiation of the random variables in Algorithm~\ref{algos}.
Hence by the same reasoning as in Claim~\ref{claim:tau-guarded} the structure $\fB$ is $\TTT^*$-guarded.  
If we show that $\fB$ also satisfies the $\TTT^*$-extension property, then Lemma~\ref{lemma:witness-of-satisfiability} will imply that $\fB \models \phi$.
The main technical step is Claim~\ref{claim:inverse}, being a deterministic replacement for Claim~\ref{claim:tau-extension-a}.
Then Claim~\ref{claim:det-tau-extension} proves that all required witnesses indeed exist in the model $\fB$.

\begin{clm}\label{claim:inverse}
  Let $k \in [0,\Width(\sigma)-1]$, and let $\tau \in \TTT^*_{k+1}$ be a $(k{+}1)$-type.  
  Consider a tuple
  \(
    \langle (\alpha_1,\beta_1),\dots,(\alpha_k,\beta_k) \rangle \in \Omega_k
  \)
  together with an index $\alpha_{k+1} \in [0,\Width(\sigma)-1] \setminus \{ \alpha_1,\dots,\alpha_k \}$.
  Then there exists some $\beta_{k+1} \in [0,M-1]$ such that the following holds:  
  for every $t \in [0,k]$ and every sequence $1 \le j_1 < \dots < j_t \le k$,
  \[
    \cX\big((\alpha_{j_1},\beta_{j_1}),\dots,(\alpha_{j_t},\beta_{j_t}),(\alpha_{k+1},\beta_{k+1})\big)
      = \type{\tau}{x_{j_1},\dots,x_{j_t},x_{k+1}}.
  \]
\end{clm}
\begin{proof}
  To prove the claim, we need to find $\beta_{k+1} \in [0,M-1]$ such that for every subset $S \subseteq [0,k+1]$ with $k+1 \in S$ the following holds.

  Let $m = |S|$, and let $r_1,\dots,r_m$ enumerate $S$ so that $\alpha_{r_1} < \dots < \alpha_{r_m}$.
  The hash value computed at Line~\ref{det-line:hash} when processing $\langle (\alpha_{r_1},\beta_{r_1}),\dots,(\alpha_{r_m},\beta_{r_m}) \rangle \in \Omega_m$ shall satisfy:
  \begin{equation}\label{eqn:chinese}
    \big(\big(\beta_{r_1} + \dots + \beta_{r_m}\big) \bmod p_S\big) \bmod |\TTT^*_m| = h_S,
  \end{equation}
  where
  $p_S \eqdef p_i$ for the index $i$ as computed at Line~\ref{det-line:index}, i.e.,
  \( i = 2^{\alpha_{r_1}} + \dots + 2^{\alpha_{r_m}}; \)
  and
  $h_S$ is the position of the $m$-type $\type{\tau}{x_{r_1},\dots,x_{r_m}}$ in the enumeration of $\TTT^*_m$ as specified at Line~\ref{det-line:enum},
  i.e., $\TTT^*_m(h_S) = \type{\tau}{x_{r_1},\dots,x_{r_m}}$.
  Note that $\beta_{k+1}$ occurs in~\eqref{eqn:chinese}, as $k+1 \in S$.

  Since $h_S < |\TTT^*_m| \le |\TTT^{\sigma}_{\Width(\sigma)}| \le p_S$,
  the solvability of~\eqref{eqn:chinese} reduces to
  \begin{equation}\label{eqn:chinese2}
    \beta_{r_1} + \dots + \beta_{r_m} \equiv h_S \mod p_S.
  \end{equation}

  Now, observe a natural correspondence: 
  the subset $\{ \alpha_{r_1},\dots,\alpha_{r_m} \} \subseteq [0,\Width(\sigma)-1]$
  is encoded in the binary digits of the index $i = 2^{\alpha_{r_1}} + \dots + 2^{\alpha_{r_m}}$.

  In consequence, the moduli of~\eqref{eqn:chinese2} are distinct primes for distinct subsets $S$.
  By the Chinese remainder theorem, the set of congruences specified by~\eqref{eqn:chinese2} admits a solution $\beta_{k+1} \in [0,M-1]$, thereby proving the claim.
\end{proof}

\begin{clm}\label{claim:det-tau-extension}
  The structure $\fB$ satisfies the $\TTT^{*}$-extension property.
\end{clm}
\begin{proof}
  Since all types in~$\TTT^{*}$ are guarded and the set~$\TTT^{*}$ is closed, the $\TTT^{*}$-extension property needs to be verified only for tuples from the sets~$\Omega_k$ considered by the algorithm.

  Let $k \in [0,\Width(\sigma){-}1]$.
  Consider a tuple \(\bs = \big\langle b_1,\dots,b_k \big\rangle \in \Omega_k \),
  and write $(\alpha_{i},\beta_{i})$ for the element $b_{i}$.
  Fix a $k$-type $\tau_1 \in \TTT^{*}_k$, and a $(k{+}1)$-type $\tau_2 \in \TTT^{*}_{k+1}$ with $\tau_2 \models \tau_1$.
  Assuming that $\type{\fB}{\bs} = \tau_1$, we prove that there exists $b_{k+1} \in B \setminus (\bs \cup \Cons)$ such that $\type{\fB}{\bs,b_{k+1}} = \tau_2$.

  Arbitrarily choose $\alpha_{k+1} \in [0,\Width(\sigma)-1] \setminus \{ \alpha_1,\dots,\alpha_k \}$.
  By Claim~\ref{claim:inverse}, there exists $\beta_{k+1} \in [0,M-1]$ such that, for every $t \in [0,k]$ and every $1 \le j_1 < \dots < j_t \le k$,
  we have
  \begin{equation}\label{eqn:det-variable1}
    \cX\big((\alpha_{j_1},\beta_{j_1}),\dots,(\alpha_{j_t},\beta_{j_t}),(\alpha_{k+1},\beta_{k+1})\big) = \type{\tau_2}{x_{j_1},\dots,x_{j_t},x_{k+1}}.
  \end{equation}

  Write $b_{k+1}$ for the new element $(\alpha_{k+1},\beta_{k+1})$.
  To show that $\type{\fB}{\bs,b_{k+1}} = \tau_2$,
  we prove that, for every $m \in [k+1]$ and every $1 \le r_1 < \dots < r_m \le k+1$, it holds
  \begin{equation}\label{eqn:det-variable2}
    \type{\str{B}}{b_{r_1},\dots,b_{r_m}} = \type{\tau_2}{x_{r_1},\dots,x_{r_m}}.
  \end{equation}
  This follows by a straightforward induction over the length $m$ of subsequences $r_1 < \dots < r_m$:
  if $r_m \le k$, then $\type{\str{B}}{b_{r_1},\dots,b_{r_m}}$ was assumed to be set consistently with $\tau_1$, and since $\tau_2 \models \tau_1$, this agrees with $\tau_2$.
  If $r_m = k + 1$, then the choice of $b_{k+1}$ (equation~\eqref{eqn:det-variable1}) guarantees us that the value of $\cX(b_{r_1},\dots,b_{r_m})$ is exactly $\type{\tau_2}{x_{r_1},\dots,x_{r_m}}$.
  By inductive hypothesis,
  the $m$-types $\type{\str{B}}{b_{r_1},\dots,b_{r_m}}$ and $\cX(b_{r_1},\dots,b_{r_m})$ are compatible (the test at Line~\ref{det-line:test} succeeds),
  and hence the algorithm assigns the $m$-type $\cX(b_{r_1},\dots,b_{r_m})$ induced from $\tau_2$, as claimed.
\end{proof}

\begin{rem}
\label{remark:dense-types} 
In general witnesses of satisfiability are allowed to contain non-guarded types. 
However, we have the following reduction ensuring that all types are guarded. 
Let $\sigma$ be the original signature and let $\TTT^* = \bigcup_{k=0}^{\Width(\sigma)} \TTT^*_k$ be the original witness of satisfiability.  
Define $\tilde{\sigma}$ as the extension of $\sigma$ with a fresh relation symbol $G$ of arity $\Width(\sigma)$.  
Each $k$-type $\tau \in \TTT^*_k$ is then modified by adding the atoms $G(\ys)$ for every sequence $\ys \in (\{x_1,\dots,x_k\} \cup \Cons)^{\Width(\sigma)}$.  
Denote by $\tilde{\TTT}^* = \bigcup_{k=0}^{\Width(\sigma)} \tilde{\TTT}^*_k$ the resulting witness of satisfiability.
It is straightforward to verify that this reduction is sound.
Moreover, it does not increase the number of $k$-types in the witness of satisfiability.
In particular $|\tilde{\TTT}^*_k| \le |\TTT^{\sigma}_k|$ holds for the original signature $\sigma$, preserving validity of size estimations in terms of $|\TTT^{\sigma}_k|$.
\end{rem}

\subparagraph{Hash Parameters.}
To derive Proposition~\ref{prop:det-fmp-normal-form} from Lemma~\ref{lemma:det-algo-analysis},
it remains to show that suitable hash parameters $p_1 < \dots < p_{2^{\Width(\sigma)}-1}$ can indeed be chosen.

A simple approach is to invoke the classical Bertrand's postulate (see: Chapter 2 in~\cite{aigner2018proofs}): for every integer $m \ge 2$, there exists a prime $p$ with $m < p < 2 \cdot m$.
We use this postulate to establish the desired bound on $\Width(\sigma) \cdot M$, i.e., the size of the unnamed domain of $\fB$.

\begin{clm}\label{claim:bertrand-bound}
  There exists a fixed constant $C>0$ such that the following holds.
  For every signature $\sigma$, there exist prime numbers $p_1,\dots,p_{2^{\Width(\sigma)}-1}$ satisfying
  \begin{enumerate}
    \item $|\TTT^{\sigma}_{\Width(\sigma)}| \le p_1  < \dots < p_{2^{\Width(\sigma)}-1}$, and
    \item $\Width(\sigma) \cdot M \le C \cdot |\TTT^{\sigma}_{\Width(\sigma)}|^{2^{\Width(\sigma)}}$, where $M = \prod_{i=1}^{2^{\Width(\sigma)}-1} p_i$.
  \end{enumerate}
\end{clm}
\begin{proof}
Starting with $|\TTT^{\sigma}_{\Width(\sigma)}|$ and applying Bertrand's postulate iteratively $2^{\Width(\sigma)}-1$ times,
we obtain distinct prime numbers $p_1,\dots,p_{2^{\Width(\sigma)}-1}$ with \( 2^{i-1} \cdot |\TTT^{\sigma}_{\Width(\sigma)}| < p_i < 2^{i} \cdot |\TTT^{\sigma}_{\Width(\sigma)}| \) for all $i \in [2^{\Width(\sigma)}-1]$.
This suffices to establish the following bound on the number $M$:
\begin{equation}\label{eqn:bertrand-estimate}
  \prod_{i=1}^{2^{\Width(\sigma)}-1} p_i
  \le 2^{1+\dots+2^{\Width(\sigma)}-1} \cdot |\TTT^{\sigma}_{\Width(\sigma)}|^{2^{\Width(\sigma)}-1}
  \le 2^{4^{\Width(\sigma)}/2} \cdot |\TTT^{\sigma}_{\Width(\sigma)}|^{2^{\Width(\sigma)}-1}.
\end{equation}

We now analyse two cases: $\Width(\sigma) \ge 4$ and $\Width(\sigma) \le 3$.
First suppose $\Width(\sigma) \ge 4$.
Then it holds $4^{\Width(\sigma)} \le \Width(\sigma)^{\Width(\sigma)}$.
From Claim~\ref{claim:number-of-types}, we have $2^{\Width(\sigma)^{\Width(\sigma)}} \le |\TTT^{\sigma}_{\Width(\sigma)}|$.
Therefore:
\begin{equation}
  M \le 2^{4^{\Width(\sigma)}/2} \cdot |\TTT^{\sigma}_{\Width(\sigma)}|^{2^{\Width(\sigma)}-1} \le 
  2^{\Width(\sigma)^{\Width(\sigma)}/2} \cdot |\TTT^{\sigma}_{\Width(\sigma)}|^{2^{\Width(\sigma)}-1} \le 
  |\TTT^{\sigma}_{\Width(\sigma)}|^{2^{\Width(\sigma)} - 1/2}.
\end{equation}
Consequently, we obtain the following bound on $\Width(\sigma) \cdot M$:
\begin{equation}\label{eqn:bertrand-large-width}
  \Width(\sigma) \cdot |\TTT^{\sigma}_{\Width(\sigma)}|^{2^{\Width(\sigma)} - 1/2}
  \le |\TTT^{\sigma}_{\Width(\sigma)}|^{1/2} \cdot |\TTT^{\sigma}_{\Width(\sigma)}|^{2^{\Width(\sigma)} - 1/2}
  \le |\TTT^{\sigma}_{\Width(\sigma)}|^{2^{\Width(\sigma)}}.
\end{equation}

Inequality~\eqref{eqn:bertrand-large-width} proves the claim whenever the width of the signature is at least $4$.
Yet below this threshold,
we may directly use~\eqref{eqn:bertrand-estimate}:
in this case
$\Width(\sigma)\cdot M$ becomes bounded by $\Width(\sigma) \cdot 2^{4^{\Width(\sigma)}/2} \cdot |\TTT^{\sigma}_{\Width(\sigma)}|^{2^{\Width(\sigma)}-1} \le C \cdot |\TTT^{\sigma}_{\Width(\sigma)}|^{2^{\Width(\sigma)}}$ for a constant $C = 3 \cdot 2^{32}$.
\end{proof}

\section{Conclusions}
\label{sec:conclusions}

In this work, we presented a new probabilistic proof of the finite model property for the Guarded Fragment.  
Our methods yield tight bounds on the size of minimal models and extend naturally to the Triguarded Fragment.  
To the best of our knowledge, no previous work on the Guarded Fragment has employed a similar probabilistic technique.  

Several natural directions remain open.
A particularly appealing challenge is to extend the applicability of probabilistic methods to stronger fragments.  
One candidate is the Clique Guarded Fragment (\CGF{}), which is known to enjoy the finite model property~\cite{BGO14}.  
Here, however, a direct probabilistic approach seems out of reach.  
For instance, in a graph-theoretic setting with a single relation $E$,  
\CGF{} can express the absence of triangles via the sentence  
\[
  \forall x,y,z\; \big(\big(E(x,y) \wedge E(y,z) \wedge E(z,x)\big) \rightarrow \bot\big).
\]  
Yet if edges were placed obliviously at random, a triangle would appear almost with certainty.
This suggests that some additional underlying structure is needed, and that future work may profit from combining probabilistic reasoning with classical model constructions.  

\section*{Acknowledgements}
  The author is grateful to Prof.~Emanuel Kiero\'nski for his guidance and supervision throughout this work.  
  The author also thanks the third reviewer of ICALP~2025 for providing detailed and valuable feedback on an earlier version of this paper.
  This article is a full version of a STACS~2026 conference paper~\cite{Fiuk26a}.

\bibliographystyle{alphaurl}
\bibliography{bib2doi}

@book{aigner2018proofs,
  title = {Proofs from {THE} {BOOK} {(3.} ed.)},
  author = {Martin Aigner and G{\"{u}}nter M. Ziegler},
  year = {2004},
  publisher = {Springer},
  doi = {10.1007/978-3-662-57265-8},
  timestamp = {Mon, 18 Jan 2016 00:00:00 +0100},
  biburl = {https://dblp.org/rec/books/daglib/0019107.bib},
  bibsource = {dblp computer science bibliography, https://dblp.org},
  isbn = {978-3-540-40460-6},
  _bib2doi_selected = {dblp:/rec/books/daglib/0019107.bib},
  _bib2doi_confirmed = {true},
  _bib2doi_finished = {true},
}

@article{ABN98,
  author = {Hajnal Andr{\'{e}}ka and Istv{\'{a}}n N{\'{e}}meti and Johan van Benthem},
  title = {Modal Languages and Bounded Fragments of Predicate Logic},
  journal = {J. Philos. Log.},
  volume = {27},
  pages = {217--274},
  year = {1998},
  timestamp = {Mon, 11 May 2020 01:00:00 +0200},
  biburl = {https://dblp.org/rec/journals/jphil/AndrekaNB98.bib},
  bibsource = {dblp computer science bibliography, https://dblp.org},
  doi = {10.1023/A:1004275029985},
  number = {3},
  url = {https://doi.org/10.1023/A:1004275029985},
  _bib2doi_selected = {dblp:/rec/journals/jphil/AndrekaNB98.bib},
  _bib2doi_confirmed = {true},
  _bib2doi_finished = {true},
}

@book{Baader_Horrocks_Lutz_Sattler_2017,
  place={Cambridge},
  title={An Introduction to Description Logic},
  publisher={Cambridge University Press},
  author={Franz Baader and Ian Horrocks and Carsten Lutz and Uli Sattler},
  year={2017},
  isbn = {9781139025355},
  doi = {10.1017/9781139025355},
}

@article{BGO14,
  author = {Vince B{\'{a}}r{\'{a}}ny and Georg Gottlob and Martin Otto},
  title = {Querying the Guarded Fragment},
  journal = {Log. Methods Comput. Sci.},
  volume = {10},
  number = {2},
  year = {2014},
  timestamp = {Thu, 25 Jun 2020 01:00:00 +0200},
  biburl = {https://dblp.org/rec/journals/corr/BaranyGO13.bib},
  bibsource = {dblp computer science bibliography, https://dblp.org},
  doi = {10.2168/LMCS-10(2:3)2014},
  url = {https://doi.org/10.2168/LMCS-10(2:3)2014},
  _bib2doi_selected = {dblp:/rec/journals/corr/BaranyGO13.bib},
  _bib2doi_confirmed = {true},
  _bib2doi_finished = {true},
}

@article{BCO12,
  author = {Vince B{\'{a}}r{\'{a}}ny and Balder ten Cate and Martin Otto},
  title = {Queries with Guarded Negation},
  journal = {Proc. {VLDB} Endow.},
  volume = {5},
  number = {11},
  pages = {1328--1339},
  year = {2012},
  timestamp = {Sat, 25 Apr 2020 01:00:00 +0200},
  biburl = {https://dblp.org/rec/journals/pvldb/BaranyCO12.bib},
  bibsource = {dblp computer science bibliography, https://dblp.org},
  doi = {10.14778/2350229.2350250},
  url = {http://vldb.org/pvldb/vol5/p1328\_vincebarany\_vldb2012.pdf},
  _bib2doi_selected = {dblp:/rec/journals/pvldb/BaranyCO12.bib},
  _bib2doi_confirmed = {true},
  _bib2doi_finished = {true},
}

@article{BtCS15,
  author = {Vince B{\'{a}}r{\'{a}}ny and Balder ten Cate and Luc Segoufin},
  title = {Guarded Negation},
  journal = {J. {ACM}},
  volume = {62},
  number = {3},
  pages = {22:1--22:26},
  year = {2015},
  timestamp = {Tue, 06 Nov 2018 00:00:00 +0100},
  biburl = {https://dblp.org/rec/journals/jacm/BaranyCS15.bib},
  bibsource = {dblp computer science bibliography, https://dblp.org},
  doi = {10.1145/2701414},
  url = {https://doi.org/10.1145/2701414},
  _bib2doi_selected = {dblp:/rec/journals/jacm/BaranyCS15.bib},
  _bib2doi_confirmed = {true},
  _bib2doi_finished = {true},
}

@book{BGG,
  author = {Egon B{\"{o}}rger and Erich Gr{\"{a}}del and Yuri Gurevich},
  title = {The Classical Decision Problem},
  publisher = {Springer},
  series = {Perspectives in Mathematical Logic},
  year = {1997},
  timestamp = {Mon, 06 Nov 2006 00:00:00 +0100},
  biburl = {https://dblp.org/rec/books/sp/BorgerGG1997.bib},
  bibsource = {dblp computer science bibliography, https://dblp.org},
  _bib2doi_selected = {dblp:/rec/books/sp/BorgerGG1997.bib},
  _bib2doi_confirmed = {true},
  _bib2doi_finished = {true},
}

@inproceedings{BMP17,
  author = {Pierre Bourhis and Michael Morak and Andreas Pieris},
  title = {Making Cross Products and Guarded Ontology Languages Compatible},
  booktitle = {Proceedings of the Twenty-Sixth International Joint Conference on Artificial Intelligence, {IJCAI} 2017, Melbourne, Australia, August 19-25, 2017},
  pages = {880--886},
  year = {2017},
  timestamp = {Tue, 20 Aug 2019 01:00:00 +0200},
  biburl = {https://dblp.org/rec/conf/ijcai/BourhisMP17.bib},
  bibsource = {dblp computer science bibliography, https://dblp.org},
  doi = {10.24963/ijcai.2017/122},
  publisher = {ijcai.org},
  url = {https://doi.org/10.24963/ijcai.2017/122},
  editor = {Carles Sierra},
  _bib2doi_selected = {dblp:/rec/conf/ijcai/BourhisMP17.bib},
  _bib2doi_confirmed = {true},
  _bib2doi_finished = {true},
}

@book{cover2006elements,
  author = {Thomas M. Cover and Joy A. Thomas},
  title = {Elements of information theory {(2.} ed.)},
  year = {2006},
  publisher = {Wiley},
  timestamp = {Wed, 10 Jul 2019 01:00:00 +0200},
  biburl = {https://dblp.org/rec/books/daglib/0016881.bib},
  bibsource = {dblp computer science bibliography, https://dblp.org},
  isbn = {978-0-471-24195-9},
  _bib2doi_selected = {dblp:/rec/books/daglib/0016881.bib},
  _bib2doi_confirmed = {true},
  _bib2doi_finished = {true},
}

@article{Fagin76,
  issn = {00224812},
  author = {Ronald Fagin},
  journal = {J. Symb. Log.},
  number = {1},
  pages = {50--58},
  title = {Probabilities on Finite Models},
  volume = {41},
  year = {1976},
  timestamp = {Wed, 14 Nov 2018 00:00:00 +0100},
  biburl = {https://dblp.org/rec/journals/jsyml/Fagin76.bib},
  bibsource = {dblp computer science bibliography, https://dblp.org},
  doi = {10.1017/S0022481200051756},
  url = {https://doi.org/10.1017/S0022481200051756},
  _bib2doi_selected = {dblp:/rec/journals/jsyml/Fagin76.bib},
  _bib2doi_confirmed = {true},
  _bib2doi_finished = {true},
}

@article{FK-lmcs25,
  title = {Alternating Quantifiers in Uniform One-Dimensional Fragments with an Excursion into Three-Variable Logic},
  author = {Oskar Fiuk and Emanuel Kieronski},
  doi = {10.46298/lmcs-21(1:25)2025},
  journal = {Log. Methods Comput. Sci.},
  issn = {1860-5974},
  volume = {21},
  eid = {25},
  year = {2025},
  month = {Mar},
  keywords = {Computer Science - Logic in Computer Science},
  timestamp = {Wed, 02 Apr 2025 01:00:00 +0200},
  biburl = {https://dblp.org/rec/journals/lmcs/FiukK25.bib},
  bibsource = {dblp computer science bibliography, https://dblp.org},
  number = {1},
  url = {https://doi.org/10.46298/lmcs-21(1:25)2025},
  _bib2doi_selected = {dblp:/rec/journals/lmcs/FiukK25.bib},
  _bib2doi_confirmed = {true},
  _bib2doi_finished = {true},
}

@inproceedings{FKM24,
  author = {Oskar Fiuk and Emanuel Kieronski and Vincent Michielini},
  title = {On the complexity of Maslov's class {K}},
  booktitle = {Proceedings of the 39th Annual {ACM/IEEE} Symposium on Logic in Computer Science, {LICS} 2024, Tallinn, Estonia, July 8-11, 2024},
  pages = {35:1--35:14},
  doi = {10.1145/3661814.3662097},
  year = {2024},
  timestamp = {Sun, 19 Jan 2025 00:00:00 +0100},
  biburl = {https://dblp.org/rec/conf/lics/FiukKM24.bib},
  bibsource = {dblp computer science bibliography, https://dblp.org},
  publisher = {{ACM}},
  url = {https://doi.org/10.1145/3661814.3662097},
  editor = {Pawel Sobocinski and Ugo Dal Lago and Javier Esparza},
  _bib2doi_selected = {dblp:/rec/conf/lics/FiukKM24.bib},
  _bib2doi_confirmed = {true},
  _bib2doi_finished = {true},
}

@article{Gol84,
  author = {Warren D. Goldfarb},
  title = {The Unsolvability of the Godel Class with Identity},
  journal = {J. Symb. Log.},
  volume = {49},
  year = {1984},
  pages = {1237--1252},
  doi = {10.2307/2274274},
  timestamp = {Sun, 28 May 2017 01:00:00 +0200},
  biburl = {https://dblp.org/rec/journals/jsyml/Goldfarb84.bib},
  bibsource = {dblp computer science bibliography, https://dblp.org},
  number = {4},
  url = {https://doi.org/10.2307/2274274},
  _bib2doi_selected = {dblp:/rec/journals/jsyml/Goldfarb84.bib},
  _bib2doi_confirmed = {true},
  _bib2doi_finished = {true},
}

@article{Gol89,
  issn = {00224812},
  author = {Warren D. Goldfarb},
  journal = {J. Symb. Log.},
  number = {2},
  pages = {460--466},
  title = {Random Models and the Maslov Class},
  volume = {54},
  year = {1989},
  timestamp = {Sun, 28 May 2017 01:00:00 +0200},
  biburl = {https://dblp.org/rec/journals/jsyml/Goldfarb89.bib},
  bibsource = {dblp computer science bibliography, https://dblp.org},
  doi = {10.2307/2274860},
  url = {https://doi.org/10.2307/2274860},
  _bib2doi_selected = {dblp:/rec/journals/jsyml/Goldfarb89.bib},
  _bib2doi_confirmed = {true},
  _bib2doi_finished = {true},
}

@article{Gol93,
  author = {Warren D. Goldfarb},
  title = {Random Models and Solvable Skolem Classes},
  journal = {J. Symb. Log.},
  volume = {58},
  year = {1993},
  pages = {908--914},
  doi = {10.2307/2275103},
  timestamp = {Sun, 28 May 2017 01:00:00 +0200},
  biburl = {https://dblp.org/rec/journals/jsyml/Goldfarb93.bib},
  bibsource = {dblp computer science bibliography, https://dblp.org},
  number = {3},
  url = {https://doi.org/10.2307/2275103},
  _bib2doi_selected = {dblp:/rec/journals/jsyml/Goldfarb93.bib},
  _bib2doi_confirmed = {true},
  _bib2doi_finished = {true},
}

@article{GGS84,
  title = {A Decidable Subclass of the Minimal Godel Class with Identity},
  volume = {49},
  doi = {10.2307/2274275},
  number = {4},
  journal = {J. Symb. Log.},
  author = {Warren D. Goldfarb and Yuri Gurevich and Saharon Shelah},
  year = {1984},
  pages = {1253--1261},
  timestamp = {Sat, 05 Sep 2020 01:00:00 +0200},
  biburl = {https://dblp.org/rec/journals/jsyml/GoldfarbGS84.bib},
  bibsource = {dblp computer science bibliography, https://dblp.org},
  url = {https://doi.org/10.2307/2274275},
  _bib2doi_selected = {dblp:/rec/journals/jsyml/GoldfarbGS84.bib},
  _bib2doi_confirmed = {true},
  _bib2doi_finished = {true},
}

@inproceedings{Gra99b,
  author = {Erich Gr{\"{a}}del},
  title = {Invited Talk: Decision procedures for guarded logics},
  booktitle = {Automated Deduction - CADE-16, 16th International Conference on Automated Deduction, Trento, Italy, July 7-10, 1999, Proceedings},
  series = {Lecture Notes in Computer Science},
  volume = {1632},
  pages = {31--51},
  publisher = {Springer},
  year = {1999},
  doi = {10.1007/3-540-48660-7_3},
  timestamp = {Sun, 21 May 2017 01:00:00 +0200},
  biburl = {https://dblp.org/rec/conf/cade/Gradel99.bib},
  bibsource = {dblp computer science bibliography, https://dblp.org},
  editor = {Harald Ganzinger},
  _bib2doi_old_doi = {10.1007/3-540-48660-7\_3},
  _bib2doi_selected = {dblp:/rec/conf/cade/Gradel99.bib},
  _bib2doi_confirmed = {true},
  _bib2doi_finished = {true},
}

@article{Gra99,
  author = {Erich Gr{\"{a}}del},
  title = {On The Restraining Power of Guards},
  journal = {J. Symb. Log.},
  volume = {64},
  number = {4},
  year = {1999},
  pages = {1719--1742},
  timestamp = {Sun, 28 May 2017 01:00:00 +0200},
  biburl = {https://dblp.org/rec/journals/jsyml/Gradel99.bib},
  bibsource = {dblp computer science bibliography, https://dblp.org},
  doi = {10.2307/2586808},
  url = {https://doi.org/10.2307/2586808},
  _bib2doi_selected = {dblp:/rec/journals/jsyml/Gradel99.bib},
  _bib2doi_confirmed = {true},
  _bib2doi_finished = {true},
}

@article{GKV97,
  author = {Erich Gr{\"{a}}del and Phokion G. Kolaitis and Moshe Y. Vardi},
  title = {On the decision problem for two-variable first-order logic},
  journal = {Bull. Symb. Log.},
  volume = {3},
  number = {1},
  year = {1997},
  pages = {53--69},
  timestamp = {Fri, 03 Jul 2020 01:00:00 +0200},
  biburl = {https://dblp.org/rec/journals/bsl/GradelKV97.bib},
  bibsource = {dblp computer science bibliography, https://dblp.org},
  doi = {10.2307/421196},
  url = {https://doi.org/10.2307/421196},
  _bib2doi_selected = {dblp:/rec/journals/bsl/GradelKV97.bib},
  _bib2doi_confirmed = {true},
  _bib2doi_finished = {true},
}

@inproceedings{GW99,
  author = {Erich Gr{\"{a}}del and Igor Walukiewicz},
  title = {Guarded Fixed Point Logic},
  booktitle = {14th Annual {IEEE} Symposium on Logic in Computer Science, Trento, Italy, July 2-5, 1999},
  pages = {45--54},
  publisher = {{IEEE} Computer Society},
  year = {1999},
  url = {https://doi.org/10.1109/LICS.1999.782585},
  doi = {10.1109/LICS.1999.782585},
  timestamp = {Fri, 24 Mar 2023 00:00:00 +0100},
  biburl = {https://dblp.org/rec/conf/lics/GradelW99.bib},
  bibsource = {dblp computer science bibliography, https://dblp.org},
  _bib2doi_selected = {dblp:/rec/conf/lics/GradelW99.bib},
  _bib2doi_confirmed = {true},
  _bib2doi_finished = {true},
}

@article{GS83,
  author = {Yuri Gurevich and Saharon Shelah},
  title = {Random Models and the Godel Case of the Decision Problem},
  journal = {J. Symb. Log.},
  volume = {48},
  number = {4},
  year = {1983},
  pages = {1120--1124},
  doi = {10.2307/2273674},
  timestamp = {Sat, 05 Sep 2020 01:00:00 +0200},
  biburl = {https://dblp.org/rec/journals/jsyml/GurevichS83b.bib},
  bibsource = {dblp computer science bibliography, https://dblp.org},
  url = {https://doi.org/10.2307/2273674},
  _bib2doi_selected = {dblp:/rec/journals/jsyml/GurevichS83b.bib},
  _bib2doi_confirmed = {true},
  _bib2doi_finished = {true},
}

@article{Her95,
  author = {Bernhard Herwig},
  title = {Extending Partial Isomorphisms on Finite Structures},
  journal = {Comb.},
  volume = {15},
  year = {1995},
  timestamp = {Wed, 22 Jul 2020 01:00:00 +0200},
  biburl = {https://dblp.org/rec/journals/combinatorica/Herwig95.bib},
  bibsource = {dblp computer science bibliography, https://dblp.org},
  doi = {10.1007/BF01299742},
  number = {3},
  pages = {365--371},
  url = {https://doi.org/10.1007/BF01299742},
  _bib2doi_selected = {dblp:/rec/journals/combinatorica/Herwig95.bib},
  _bib2doi_confirmed = {true},
  _bib2doi_finished = {true},
}

@phdthesis{hirsch2002guarded,
  author = {Colin Hirsch},
  title = {Guarded logics: algorithms and bisimulation},
  type = {Doctoral dissertation},
  school = {{RWTH} Aachen University, Germany},
  location = {Aachen, Germany},
  date = {2002-07-17},
  supervisor = {Grädel, Erich and Thomas, Wolfgang},
  timestamp = {Tue, 18 Jan 2022 00:00:00 +0100},
  biburl = {https://dblp.org/rec/phd/dnb/Hirsch02.bib},
  bibsource = {dblp computer science bibliography, https://dblp.org},
  year = {2002},
  urn = {urn:nbn:de:hbz:82-opus-4550},
  _bib2doi_selected = {dblp:/rec/phd/dnb/Hirsch02.bib},
  _bib2doi_confirmed = {true},
  _bib2doi_finished = {true},
}

@phdthesis{Kaz06,
  author = {Yevgeny Kazakov},
  title = {Saturation-based decision procedures for extensions of the guarded fragment},
  school = {Universit\"at des Saarlandes},
  year = {2006},
  address = {Saarbr\"ucken, Germany},
  _bib2doi_finished = {true},
}

@inproceedings{KR21lics,
  author = {Emanuel Kieronski and Sebastian Rudolph},
  title = {Finite Model Theory of the Triguarded Fragment and Related Logics},
  booktitle = {36th Annual {ACM/IEEE} Symposium on Logic in Computer Science, {LICS} 2021, Rome, Italy, June 29 - July 2, 2021},
  pages = {1--13},
  doi = {10.1109/LICS52264.2021.9470734},
  year = {2021},
  timestamp = {Mon, 03 Jan 2022 00:00:00 +0100},
  biburl = {https://dblp.org/rec/conf/lics/KieronskiR21.bib},
  bibsource = {dblp computer science bibliography, https://dblp.org},
  publisher = {{IEEE}},
  url = {https://doi.org/10.1109/LICS52264.2021.9470734},
  _bib2doi_selected = {dblp:/rec/conf/lics/KieronskiR21.bib},
  _bib2doi_confirmed = {true},
  _bib2doi_finished = {true},
}

@article{KT18,
  author = {Emanuel Kieronski and Lidia Tendera},
  title = {Finite Satisfiability of the Two-Variable Guarded Fragment with Transitive Guards and Related Variants},
  journal = {{ACM} Trans. Comput. Log.},
  volume = {19},
  number = {2},
  year = {2018},
  pages = {8:1--8:34},
  timestamp = {Sun, 25 Jul 2021 01:00:00 +0200},
  biburl = {https://dblp.org/rec/journals/tocl/KieronskiT18.bib},
  bibsource = {dblp computer science bibliography, https://dblp.org},
  doi = {10.1145/3174805},
  url = {https://doi.org/10.1145/3174805},
  _bib2doi_selected = {dblp:/rec/journals/tocl/KieronskiT18.bib},
  _bib2doi_confirmed = {true},
  _bib2doi_finished = {true},
}

@article{Mar01,
  author = {Maarten Marx},
  title = {Tolerance Logic},
  journal = {J. Log. Lang. Inf.},
  volume = {10},
  number = {3},
  pages = {353--374},
  year = {2001},
  doi = {10.1023/A:1011207512025},
  timestamp = {Thu, 17 Sep 2020 01:00:00 +0200},
  biburl = {https://dblp.org/rec/journals/jolli/Marx01.bib},
  bibsource = {dblp computer science bibliography, https://dblp.org},
  url = {https://doi.org/10.1023/A:1011207512025},
  _bib2doi_selected = {dblp:/rec/journals/jolli/Marx01.bib},
  _bib2doi_confirmed = {true},
  _bib2doi_finished = {true},
}

@book{PH23,
    author = {Pratt-Hartmann, Ian},
    title = {Fragments of First-Order Logic},
    publisher = {Oxford University Press},
    year = {2023},
    isbn = {9780192867964},
    doi = {10.1093/oso/9780192867964.001.0001},
    url = {https://doi.org/10.1093/oso/9780192867964.001.0001},
}

@inproceedings{rosati2006finitechase,
  author = {Riccardo Rosati},
  title = {On the decidability and finite controllability of query processing in databases with incomplete information},
  year = {2006},
  isbn = {1595933182},
  publisher = {{ACM}},
  address = {New York, NY, USA},
  doi = {10.1145/1142351.1142404},
  booktitle = {Proceedings of the Twenty-Fifth {ACM} {SIGACT-SIGMOD-SIGART} Symposium on Principles of Database Systems, June 26-28, 2006, Chicago, Illinois, {USA}},
  pages = {356--365},
  numpages = {10},
  location = {Chicago, IL, USA},
  series = {PODS '06},
  timestamp = {Thu, 02 Feb 2023 00:00:00 +0100},
  biburl = {https://dblp.org/rec/conf/pods/Rosati06.bib},
  bibsource = {dblp computer science bibliography, https://dblp.org},
  url = {https://doi.org/10.1145/1142351.1142404},
  editor = {Stijn Vansummeren},
  _bib2doi_selected = {dblp:/rec/conf/pods/Rosati06.bib},
  _bib2doi_confirmed = {true},
  _bib2doi_finished = {true},
}

@inproceedings{RS18,
  author = {Sebastian Rudolph and Mantas Simkus},
  title = {The Triguarded Fragment of First-Order Logic},
  booktitle = {{LPAR-22.} 22nd International Conference on Logic for Programming, Artificial Intelligence and Reasoning, Awassa, Ethiopia, 16-21 November 2018},
  series = {EPiC Series in Computing},
  volume = {57},
  pages = {604--619},
  year = {2018},
  timestamp = {Mon, 26 Jun 2023 01:00:00 +0200},
  biburl = {https://dblp.org/rec/conf/lpar/RudolphS18.bib},
  bibsource = {dblp computer science bibliography, https://dblp.org},
  doi = {10.29007/m8ts},
  publisher = {EasyChair},
  url = {https://doi.org/10.29007/m8ts},
  editor = {Gilles Barthe and Geoff Sutcliffe and Margus Veanes},
  _bib2doi_selected = {dblp:/rec/conf/lpar/RudolphS18.bib},
  _bib2doi_confirmed = {true},
  _bib2doi_finished = {true},
}

@techreport{Ben97,
  author = {Johan van {Benthem}},
  title = {Dynamic Bits and Pieces},
  institution = {ILLC},
  year = {1997},
  series = {LP},
  url = {https://eprints.illc.uva.nl/id/eprint/1251/},
  _bib2doi_finished = {true},
}

@article{NIVELLE200321,
title = {Deciding the guarded fragments by resolution},
journal = {J. Symb. Comput.},
volume = {35},
number = {1},
pages = {21-58},
year = {2003},
issn = {0747-7171},
doi = {10.1016/S0747-7171(02)00092-5},
author = {Hans de Nivelle and Maarten de Rijke},
}

@InProceedings{fiuk26a,
  author =	{Fiuk, Oskar},
  title =	{{Random Models and Guarded Logic}},
  booktitle =	{43rd International Symposium on Theoretical Aspects of Computer Science (STACS 2026)},
  pages =	{37:1--37:21},
  series =	{Leibniz International Proceedings in Informatics (LIPIcs)},
  ISBN =	{978-3-95977-412-3},
  ISSN =	{1868-8969},
  year =	{2026},
  volume =	{364},
  editor =	{Mahajan, Meena and Manea, Florin and McIver, Annabelle and Thang, Nguyen Kim},
  publisher =	{Schloss Dagstuhl -- Leibniz-Zentrum f{\"u}r Informatik},
  address =	{Dagstuhl, Germany},
  URL =		{https://drops.dagstuhl.de/entities/document/10.4230/LIPIcs.STACS.2026.37},
  URN =		{urn:nbn:de:0030-drops-255269},
  doi =		{10.4230/LIPIcs.STACS.2026.37}
}

\end{document}